\documentclass[twocolumn]{aastex631}

\submitjournal{ApJ}

\shorttitle{Progenitor Modeling I}
\shortauthors{Jacovich et al.}

\begin{document}

\title{A Grid of Core-Collapse Supernova Remnant Models. I: The Effect of Wind-Driven Mass-Loss}

\correspondingauthor{Taylor Jacovich}
\email{tjacovich@cfa.harvard.edu, tjacovich@gwu.edu}

\author{Taylor Jacovich}
\affil{Center for Astrophysics \textbar \ Harvard \& Smithsonian\\
60 Garden St. Cambridge, MA 02138}
\affil{The George Washington University Department of Physics\\
 725 21st Street, NW Washington, DC 20052}
 \affil{Astronomy, Physics, and Statistics Institute of Sciences (APSIS)\\ 725 21st Street NW, Washington, DC 20052}

\author{Daniel Patnaude}
\affil{Center for Astrophysics \textbar \ Harvard \& Smithsonian\\
60 Garden St. Cambridge, MA 02138}

\author{Patrick Slane}
\affil{Center for Astrophysics \textbar \ Harvard \& Smithsonian\\
60 Garden St. Cambridge, MA 02138}

\author{Carles Badenes}
\affil{Department of Physics and Astronomy and Pittsburgh Particle Physics\\ Astrophysics and Cosmology Center (PITT PACC)\\ University of Pittsburgh\\ 3941 O'Hara Street\\ Pittsburgh, PA 15260, USA}

\author{Shiu-Hang Lee}
\affil{Department of Astronomy, Kyoto University\\
Oiwake-cho, Kitashirakawa, Sakyo-ku, Kyoto 606-8502 Japan}

\author{Shigehiro Nagataki}
\affil{RIKEN\\
2-1 Hirosawa, Wako, Saitama 351-0198, Japan}

\author{Dan Milisavljevic}
\affil{Department of Physics and Astronomy\\ Purdue University, 525 Northwestern Avenue\\ West Lafayette, IN, 47907}

\begin{abstract}

Massive stars can shed material via steady, line--driven winds, eruptive outflows, or
mass-transfer onto a binary companion. 
In the case of single stars, the mass is deposited by the stellar wind into the nearby environment. After the massive star explodes, the stellar ejecta interact with this circumstellar material (CSM), often-times resulting in bright X-ray line emission from both the shock-heated CSM and ejecta.
The amount of material lost by the progenitor, the mass of ejecta,
and its energetics all impact the bulk spectral characteristics of this X-ray emission.
Here we present a grid of core-collapse supernova remnant models derived from models for massive stars with zero age main sequence masses of $\sim$ 10 -- 30 M$_{\sun}$ 
evolved from the pre-main sequence stage with wind-driven mass-loss. Evolution is handled by a multi-stage pipeline of software packages. First, we use mesa (Modules for Experiments in Stellar Astrophysics) to evolve the progenitors from pre-main sequence to iron core collapse. We then use the Supernova Explosion Code (snec) to explode the mesa models, and follow them for the first 100 days following core-collapse. Finally, we couple the snec output, along with the CSM generated from mesa mass-loss rates, into the Cosmic-Ray Hydrodynamics code (ChN) to model the remnant phase to 7000 years post core-collapse. At the end of each stage, we compare our outputs with those found in the literature, and we examine any qualitative and quantitative differences in the bulk properties of the remnants and their spectra based on the initial progenitor mass, as well as mass-loss history.

\end{abstract}


\section{Introduction}
Core-collapse (CC) supernova remnants (SNR) are important drivers of galactic evolution, as well as tracers of massive star populations and their evolution. Broadband observations have yielded a wealth of information about remnants, but coupling observations to theory requires sifting through a great deal of uncertainty in the parameters that dictate evolution \citep[e.g.,][]{patnaude17b}. In particular, stellar mass loss rates throughout the life of the progenitor star can have a strong impact on the structure and dynamics of the resulting remnant \citep{pat15,pat17}. 


Many open questions remain concerning the efficiency of both steady and episodic mass loss mechanisms, the physical conditions in the progenitor required to drive mass loss, and exactly when in the stellar life cycle certain mass loss mechanisms are more prevalent \citep[e.g.,][]{QSwave, smith14, fuller17}. In the case of some Type IIP~SNe, radio observations indicate quasi-steady mass loss up to nearly the point of core collapse \citep[e.g., SN~2011ja;][]{chakraborti16}, while optical observations of so-called "supernova impostors" suggest that some progenitors undergo massive, eruptive events only a few years before the onset of core collapse \citep[e.g., SN~2009fk; SN~2009ip;][]{moriya11,margutti14}. Thus, current theories of stellar mass loss focus on wind-driven versus episodic events following the onset of core-carbon burning \citep{pat17}. An important probe of these mass loss mechanisms comes from the forward shock emission of young SNR. The forward shock at early post-core-collapse time interacts with the CSM sculpted by late-time mass loss \citep{Ch82csm}. Forward shock evolution can therefore probe CSM density and composition at increasing radii around the SNR, and these results can yield the mass loss rates and durations as a function of age pre-core collapse \citep{Ch82mdot, CF94csm}.

While mass loss rates can depend strongly upon the mass loss mechanism, they also depend on parameters of the progenitor star such as the zero-age main-sequence (ZAMS) mass,
the stellar metallicity, rotation, and whether the star is in a binary system \citep[e.g.][]{maeda15,ouchi17}. Overall, the complexity of simply incorporating all these properties into a realistic model for the progenitor is an intractable problem. Likewise, fully realized multidimensional simulations of core collapse supernovae through and beyond shock breakout are not readily available and are often tailored to specific objects \citep[e.g.,][]{orlando20,orlando21}, nor are evolutionary models for supernova remnants in complex circumstellar environments. Finally, small refinements to model input parameters derived from observations can render a multidimensional simulation obsolete, and any one high-fidelity end to end model may not be applicable to the diverse types of core collapse supernovae and supernova remnants we observe in the local Universe. 

To address the complexities introduced when trying to develop a high fidelity model for the end-to-end evolution of a massive star from the pre-main sequence through the SNR phase, we present here a grid of one-dimensional (1D) models for core collapse SNR evolution, derived from models for massive stars. We use the "Modules for Experiments in Stellar Astrophysics" \citep[\texttt{mesa};][]{paxton1,paxton2,paxton3,paxton4,paxton5} for the stellar evolution, assuming purely line driven winds with rates from \citet{vl05} and \citet{dj88}. At the onset of core collapse, the progenitor models are coupled to the "Supernova Explosion Code" \citep[\texttt{snec};][]{moro1} modified to include explosive nucleosynthesis \citep{pat17}. Finally, the output from \texttt{snec} is coupled to our "cosmic-ray hydrodynamics" code
\citep[\texttt{ChN};][]{ellison07,ChN1,ChN2,ChN3}, to simulate the SNR evolution.

We focus on producing a wide range of progenitors such that we can cover a reasonable amount of stellar mass parameter space in \S \ref{mesa}, and calibrate our SN models such that they are in reasonable agreement with simulated compositions from the literature in \S \ref{snec}. Finally, we examine the remnants as a function of time after core-collapse with particular attention paid to X-ray spectral lines for higher Z elements (eg. S and Fe) in \S \ref{chn}.  We examine centroid information to determine the behavior of observables for the current generation of telescopes in \S \ref{ionization}. We then move to a discussion of He-like emission from high Z elements and discuss what additional power these probes may have in determining progenitor information in \S \ref{hhefs}. Additionally, we apply our models to the well-studied Galactic SNRs in \S~\ref{sec:snr_obs}. We close with a discussion of what spectral information will be available to future X-ray missions, and discuss the behavior of some observables for spatially unresolved remnants in \S \ref{unresolved}.

\section{Methods}
\label{methods}
Our current grid of models encompasses $\sim 570$ models with ZAMS masses ranging from 9.6$M_{\odot}$ to 30.0$M_{\odot}$. Each progenitor mass was modeled using various parameter settings at each stage of the pipeline, and the results are contained within publicly available git repositories for ease of versioning and to aid in reproducing our results.  In this section, we undertake a detailed description of the pipeline, as well as the various parameter choices used to generate the final models.

\subsection{Progenitor Modeling in \texttt{mesa}}
\label{mesa}
We begin our pipeline with the stellar evolution code \texttt{mesa} using version \texttt{mesa-r10398} \citep{paxton1,paxton2,paxton3,paxton4,paxton5}. Models were derived from the sample project \texttt{example\_make\_pre\_ccsne}. 
We chose to use a pre-existing set of inlists for a number of reasons, the chief ones being ease of implementation and stability across a wide range of physical parameters, as well as ease of reconstructing our pipeline against future \texttt{mesa} versions. For our baseline models, we initially left all parameters at their default, aside from $Z$ which we increased to $Z_{\odot}$. This resulted in non-rotating stars configured to use the 'Dutch' wind scheme for mass loss in the red supergiant (RSG) phase, with rates derived from \citep{dj88}.  These settings were applied to a grid with ZAMS masses ranging from 9.0$M_{\odot}$ $\lesssim$ M$_{\mathrm{ZAMS}}$ $\lesssim$ 30$M_{\odot}$. Of the 205 models, $\sim 90\%$ successfully reached core-collapse, with the remainder encountering difficulty during the Si burning phase. These failures were biased toward the lower end of the mass range ($\sim 9 \rm M_{\odot}$), with no stars below $9.6 \rm M_{\odot}$ reaching core-collapse. This is unsurprising as neutron-rich material plays a larger role in the later burning stages, and the 21-isotope network used  does not contain enough of these isotopes to accurately model these lower-mass stars \citep{paxton3}.

\begin{figure}[hbt!]
    \centering
    \includegraphics[scale=0.25]{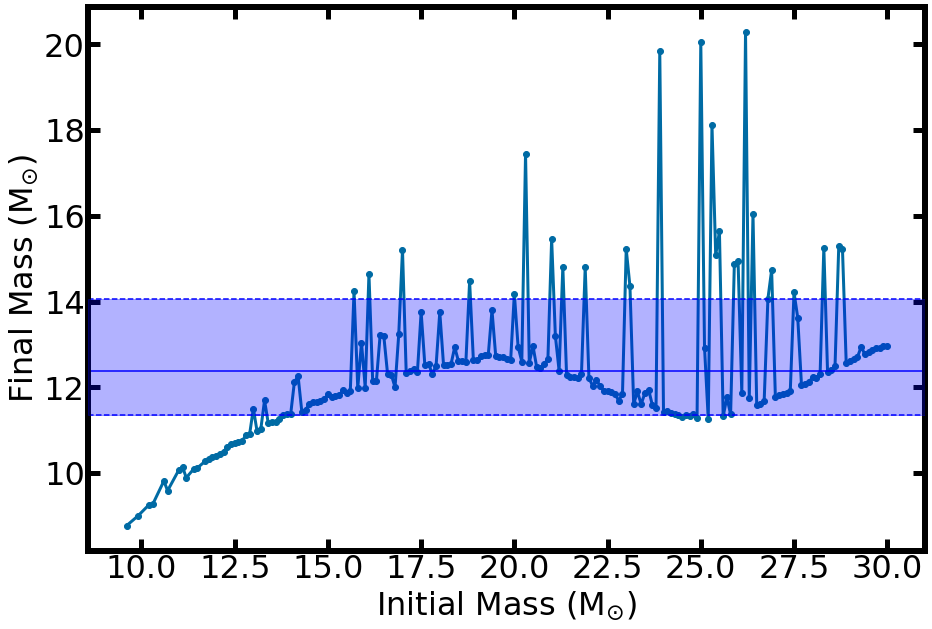}
    \caption{Final Mass vs. ZAMS mass for the rotating de Jager wind models. Shaded region corresponds to  $\bar{M}_{\rm final}=12.39^{+1.68}_{-1.04}\ \rm M_{\odot}$. Although ZAMS masses span $\sim 20\ \rm M_{\odot}$, the majority of final masses are confined to $\lesssim$ $14 \rm M_{\odot}$.}
    \label{fig:ZFM}
\end{figure}

Following our non-rotating models, we performed an additional run over the same mass range with a specified angular momentum of $\Omega=0.55\Omega_{\rm crit}$, and an initial surface velocity of $v=0.55v_{\rm crit}$. Here $v_{\rm crit}$ and $\Omega_{\rm crit}$ are defined as the maximum values of the surface velocity and total angular momentum that can be maintained by the model star before it becomes gravitationally unbound. These initial conditions were chosen to match with those used by \citet{renzo} in their explodability study.  The final masses of the models including the compact object are plotted against the ZAMS mass in Figure~\ref{fig:ZFM} for the rotating de Jager models. In spite of the large initial mass range, the final mass range encompasses a narrow set of masses, with the average final mass being $\approx 12.5 \rm M_{\odot}$. 

In examining this initial set of models, we discovered several that contained a large loop in their HR diagrams of constant luminosity, but increased effective surface temperature $(T_{\text{eff}})$, which ultimately returned to the expected late-stage evolution seen in the non-looping models.  These loops appeared to be blue-loops, as seen in the evolution of many stars \citep[e.g.][]{bl04,paxton5}. Unlike observed blue loops, the modeled stars experienced significantly reduced mass-loss due to the modeled hot winds being less effective at lifting mass from the star than the cool winds of stars that did not exhibit a loop. A closer examination showed that these stars experienced the same Fe opacity peak that would be expected for looping stars, but it did not explain why others did not experience this \citep{paxton5}.  Performing a mesh resolution study seemed to indicate that these loops are physical, as although the stars that originally exhibited the loops may have lost them at higher resolutions, the looping behavior was a constant, and ultimately settled to a narrow selection of masses.

\begin{figure}[hbt!]
    \centering
    \includegraphics[scale=0.25]{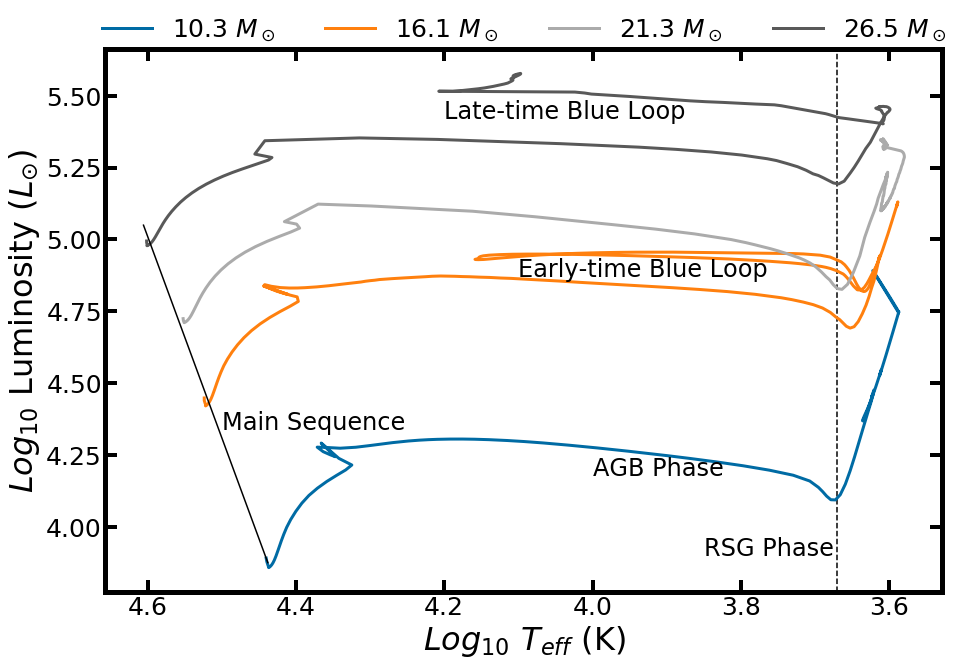}
    \caption{Example HR diagrams for four progenitor models, exhibiting the three major evolutionary tracks. We note that three of the four models explode as RSG stars, while the most luminous star at CC explodes as BSG.}
    \label{fig:HR}
\end{figure}

In addition to these models, we have two secondary grids containing models from $12-23 \rm M_{\odot}$.  One was run with a cool wind based on the van Loon wind scheme \citep{vl05}, and the other with the metallicity set to the default Z of 0.006.  Combined, these groups give us a total of 570 models simulated from pre-main-sequence to iron core-collapse. The final models were recorded along with data for several key outputs during the lifetime of the star, the chief ones being mass loss, wind velocity, temperature and luminosity.  The former two parameters are important for determining the behavior of the circumstellar environment, the latter two are important for determining where in the stellar lifetime certain mass loss events or significant changes have occurred, as well as for making sure the models are well behaved throughout the process. We present example HR diagrams for the major evolutionary tracks in Figure~\ref{fig:HR}, and discuss any variation based on the parameters. 

\begin{figure}[hbt!]
    \centering
    \includegraphics[scale=0.15]{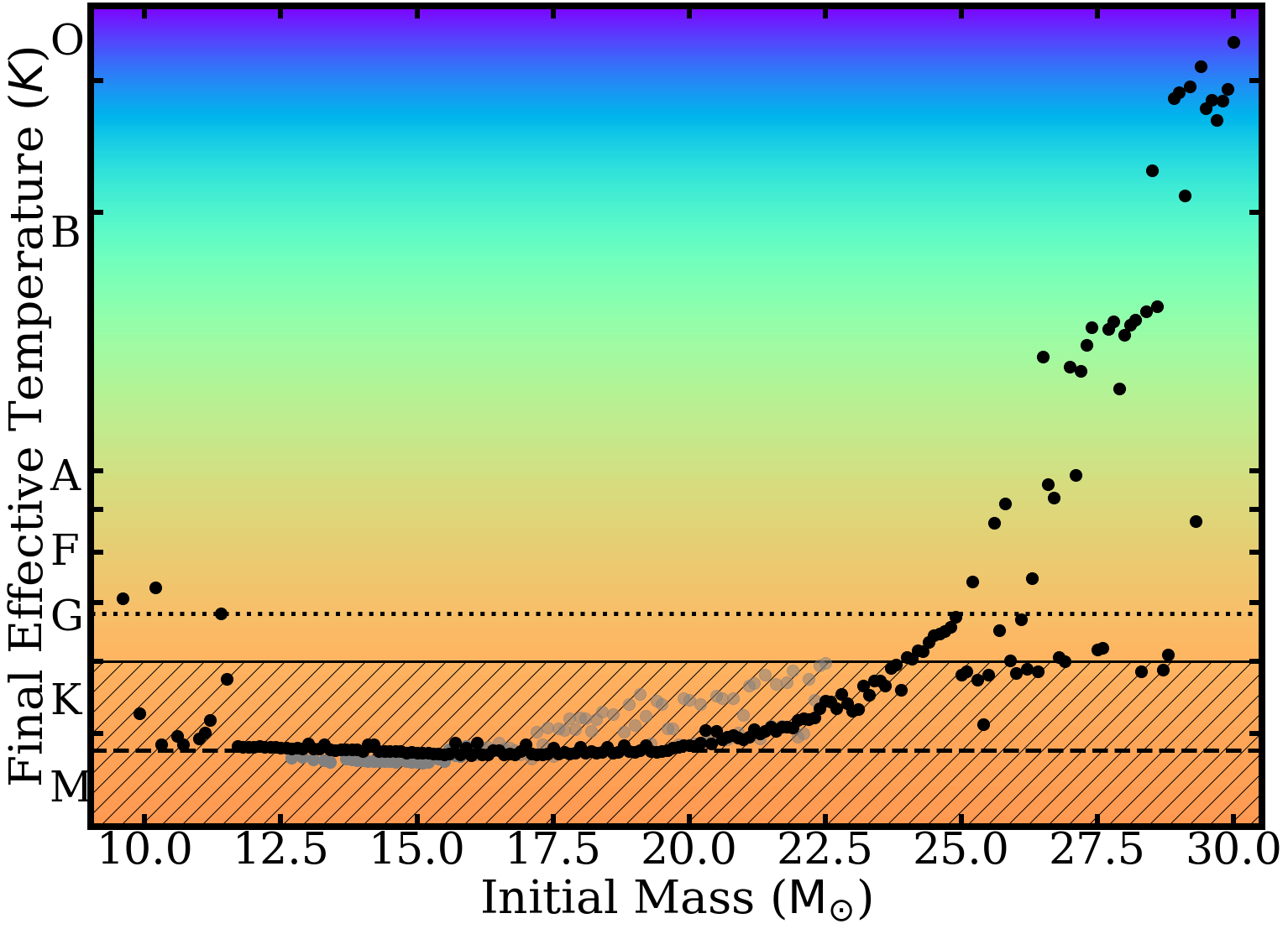}
    \caption{Final Temperature vs. ZAMS mass for the rotating models (black), and the van Loon wind models (grey). The dashed line corresponds to the hottest star in the van Loon interpolation range. The dotted line indicates solar $T_{\rm eff}$. Stars with a ZAMS mass below $\sim 23\ M_{\odot}$ result in RSG progenitors (hatched region). Above that point, stars that do not experience drops in mass-loss in the carbon-burning stages begin to explode as hotter and more compact objects.}
    \label{fig:specf}
\end{figure}


The cutoff for RSG progenitors occurs $\sim 22~ \rm M_{\odot}$.  Beyond that point, the effective temperature increases rapidly for more massive progenitors, and they quickly become yellow supergiant (YSG) and blue supergiant (BSG) stars, as can be seen in Figure \ref{fig:specf}. We note this is broadly consistent with \citet{katsuda}, who found that observed RSG progenitors are similarly limited. This is further born out by \citet{davies} who found that when considering non-RSG progenitors, along with other types of CCSNe, there is no strong indication of a mass limit on progenitors in general.

\begin{figure}[htb!]
    \centering
    \includegraphics[scale=0.42]{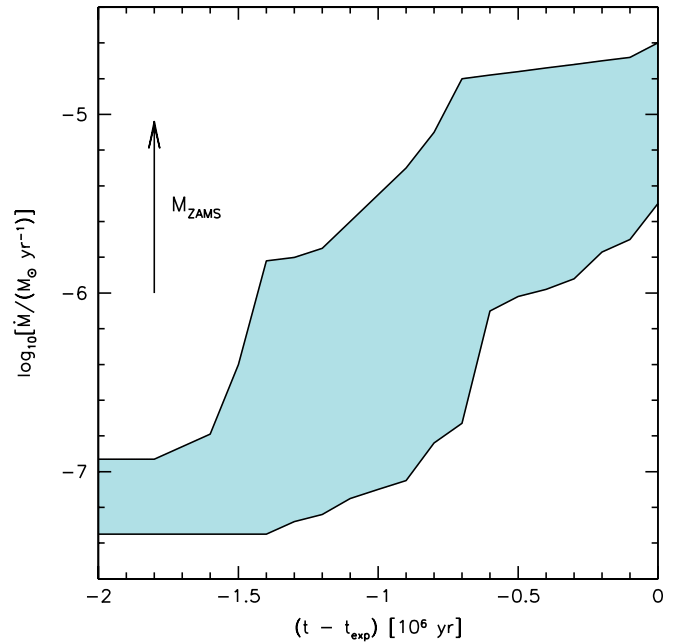}
    \caption{Mass loss rate parameter space for the rotating de Jager models. The parameter space is bounded by a $30 \rm M_{\odot}$ star above, and a $10 \rm M_{\odot}$ star below. Note how the overall trend exhibits a significant ramping up of mass loss in the last $500-800$ kyr. Higher mass stars exhibit an earlier increase around $\sim1-1.3$ Myr before CC, but the $800$ kyr peak removes nearly an order of magnitude more mass. The low mass stars exhibit very low mass-loss rates across the board, so it becomes less important to determine the exact point where mass-loss reaches RSG levels.}
    \label{fig:mass-loss}
\end{figure}

We can similarly dissect the mass-loss history of the star to determine what points would be most important for modeling the CSM that would be close enough to interact with the subsequent remnant.  In theory, we can model the entire wind from ZAMS to core-collapse, but in practice, the main-sequence winds have already reached $\gtrsim 15 \rm pc$ by core-collapse, and would thus be unlikely to interact with the remnant during the simulated timeframe. As such, we focus on the material removed from the star during the RSG phase, which generally occurs during the final 10$^5$ -- 10$^6$ years. This becomes readily apparent when examining the difference in mass-loss rates during the main sequence and early post-main sequence versus the final years of stellar life. The range of the mass-loss histories over the last two million years of evolution is presented in Figure~\ref{fig:mass-loss}.

\begin{figure}[hbt!]
    \centering
    \includegraphics[scale=0.25]{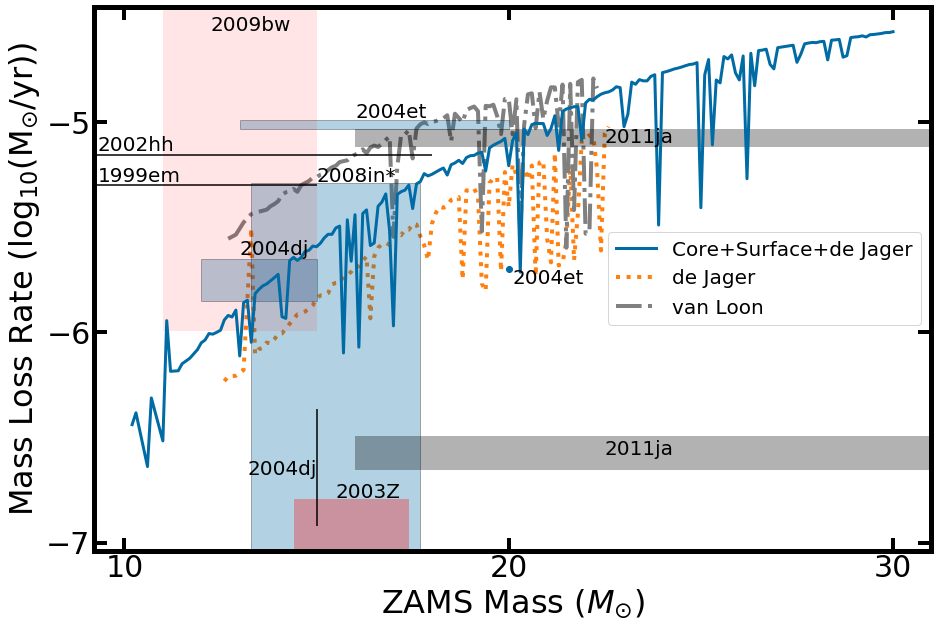}
    \caption{The average mass loss rates for the final 500 kyrs before CC as a function of ZAMS mass for the rotating de Jager models (blue line), the non-rotating de Jager models (orange line), and the non-rotating van Loon models (grey line). We compare these to derived mass-loss rates and their uncertainty regions (colored boxes) for IIp SNe as presented in \citet{dwarkadas}. Many SNe align with the model mass-loss rates.}
    \label{fig:massloss}
\end{figure}

As mentioned, many of our models reach CC as RSG stars. With this in mind, we examined the impact a dust-driven wind as described by \citet{vl05} would have on mass-loss rates. The change in mass-loss can be significant, especially for lower mass stars, as can be seen in Figure~\ref{fig:massloss}, but the applicability of the van Loon wind scheme is limited for our models. Even our coolest stars are at the upper end of the interpolation range for the wind prescription (Figure~\ref{fig:HR}), and these still spend a large fraction of their post-main-sequence life above the required temperature. Even with these caveats though, van Loon RSG winds are in reasonable agreement with the derived mass-loss rates ($\dot{M}$) for 2002hh, 1999em, and 2004et, which all have derived masses within the van Loon interpolation range of our models. At higher masses, the van Loon wind does not cover much of the post main-sequence lifetime, and the increase in mass loss is only slightly more than expected in the rotating model case.

\begin{figure}[hbt!]
    \centering
    \includegraphics[scale=0.250]{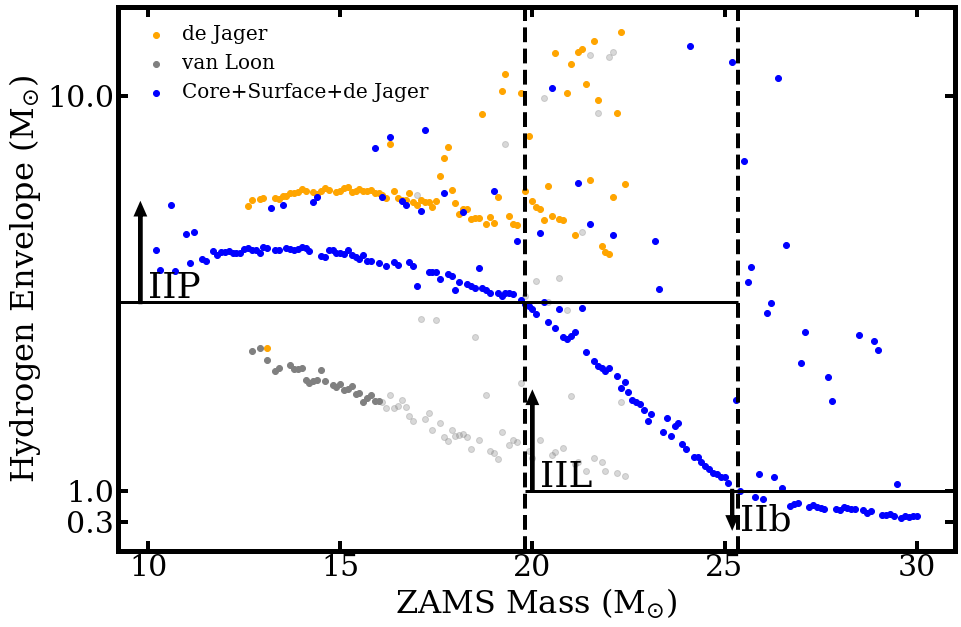}
    \caption{The Final Hydrogen Envelope mass as a function of ZAMS mass. Much like in Figure~\ref{fig:specf}, The H-envelope size is more or less constant until $\rm M\sim20 \rm M_{\odot}$, at which point the envelope mass shrinks significantly. SNe type designations are based on \citet{nomoto95}. The light grey points correspond to stars that have a $T_{\rm eff}$ above the interpolation range for the van Loon wind scheme.}
    \label{fig:henv}
\end{figure}

 As shown in Figure~\ref{fig:massloss}, all our models have wind-driven mass-loss rates of $\dot{M}\sim 10^{-7}-10^{-5}\rm M_{\odot}/yr$, which is consistent with previous studies  \citep[e.g.,][]{dwarkadas}. There is evidence of larger mass-loss rates ($\gtrsim 10^{-4}\rm M_{\odot}/yr$) for type II SNe in the final years before CC \citep[e.g.][]{forster}. The exact affect this has on the CSM will be discussed in \ref{chn}, but it would not be noticeable in our averaged mass-loss rates due to their short lifetime compared to the total post-main sequence period. The total mass loss was dictated by the chosen wind mechanism with the parameters set identically for all models. For stars at the lower end of the mass range, the RSG winds are not sufficient to remove a significant fraction of the hydrogen envelope, as seen in Figure~\ref{fig:henv}. The H envelope mass is also well correlated with the final effective temperature. Based on the H-envelope and mass-loss data, partially stripped envelope SNe (ie. IIL) do appear to be possible above $\rm M\sim 20\rm M_{\odot}$, but lower masses will need additional mass-loss mechanisms to produce these types of SNe. Stars above $\sim25 \rm M_{\odot}$ can produce IIb SNe based on \citet{nomoto95}, but Ib/c SNe will also need additional mass-loss mechanisms at all masses \citep[e.g.,][]{sravan19}.
 
\begin{figure}[hbt!]
    \centering
    \includegraphics[scale=0.25]{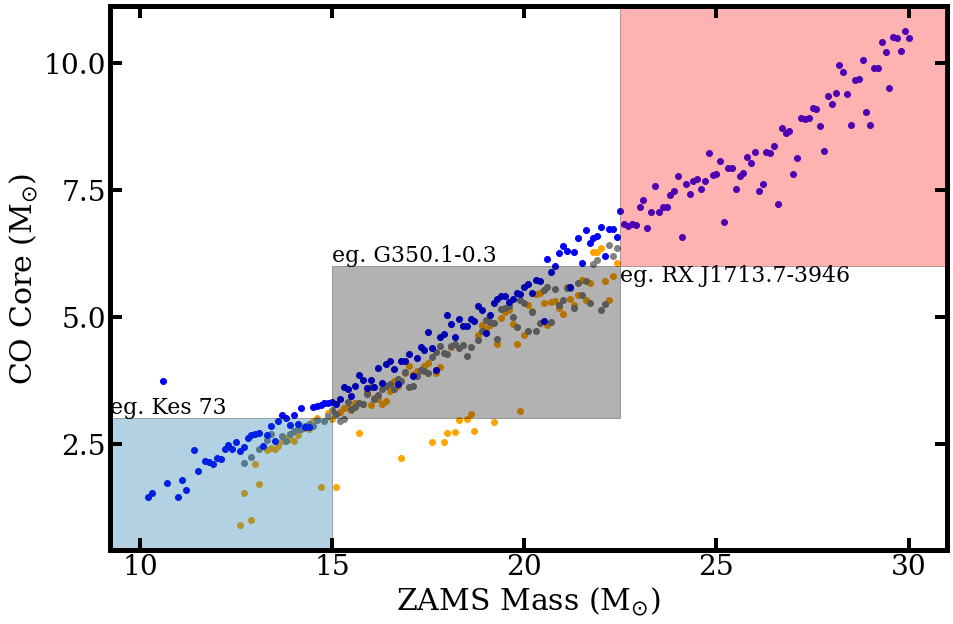}
    \caption{The Final Carbon-Oxygen Core mass as a function of ZAMS mass. The CO-core is insensitive to mass-loss, unlike many of the other metrics described. We have included example data from \citet{katsuda}, which indicate the major mass bins they determined for the CO and ZAMS masses, using the measured Fe/Si mass ratio.}
    \label{fig:COc}
\end{figure}

 \begin{figure}[hbt!]
    \centering
    \includegraphics[scale=0.35]{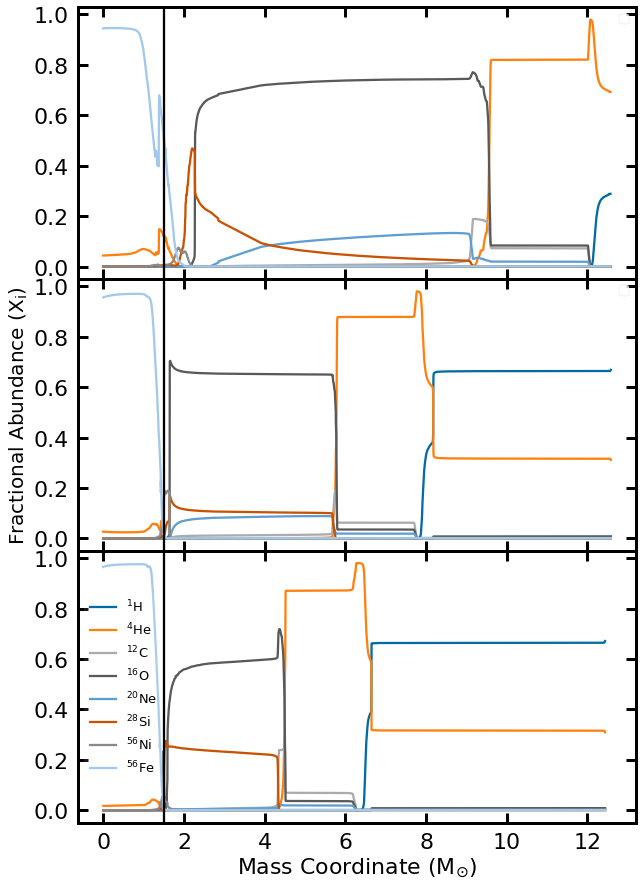}
    \caption{Stellar abundances as a function of Mass coordinate for the final mesa model for $\rm M_{\rm ZAMS} =(17.5,20.6,28.9)~\rm M_{\odot}$ (bottom to top).  These three masses were picked as they all have a final mass $\sim \bar{\rm M}$. The growth of the CO core, and the loss of the H-envelope are readily apparent. Additionally, there emerges an independent Si core in the $29 \rm M_{\odot}$ model. The peak Si abundance occurs within the CO core in the lower mass models. The vertical black line indicates the location of the excised mass for the central compact object.}
    \label{fig:final_profile}
\end{figure}

The carbon-oxygen (CO) core is strongly tied to the ZAMS mass, and seems largely insensitive to the mass-loss rates of individual models, or to the mass-loss prescription. This is particularly obvious in Figure~\ref{fig:COc} as the outliers seen in Figures~\ref{fig:massloss} and \ref{fig:henv} are absent. This points to the CO core mass being a useful proxy for determining the ZAMS mass of a remnant as stated in \citet{katsuda}. 
Figure~\ref{fig:final_profile} presents three example models with similar final masses. The growth of the CO core and the loss of the H-envelope are readily apparent. The abundance peaks are due to lighter elements burning and mixing inward at the shell boundaries. Because the final compositions are not strongly affected by the mass-loss rates, and the final mass range is quite narrow, we choose to focus our discussion on the rotating remnant models and note that non-rotating van Loon (de Jager) models largely lead to an increased (decreased) CSM density, and otherwise follow the general trends discussed below.

\subsection{Modeling Supernovae with \texttt{snec}}
\label{snec}

When the \texttt{mesa} models reach core--collapse, determined to be when Fe core infall exceeds 1000 km s$^{-1}$, 
we save the final model and pass it to the Supernova Explosion Code (\texttt{snec}). As discussed in 
\citet{pat17}, \texttt{snec} \citep{moro1} has been modified to include explosive nucleosynthesis (the so-called
\texttt{approx21} network), and a Helmholtz Equation of State \citep{timmes00}. The progenitor models are
exploded with \texttt{snec}, and evolved to an age of 100 days. The end product of the \texttt{snec} modeling is a model for the dynamics and composition of the ejecta, which can be read into \texttt{ChN} (see \S~\ref{chn}). 

Besides the input progenitor model, \texttt{snec} allows the user to vary the explosion energy, the energy injection rate, the mass over which the energy is spread, and the neutron star mass cut. By tuning these parameters, the final chemical yields can be changed in the regions of the model undergoing explosive burning. Because of this unbounded parameter space, we chose to calibrate our models against published yields from one-dimensional SNe models. In particular, we benchmarked our models against \citet[][hereafter YF07]{YF07}, though a similar calibration against other abundance sets could be explored. 


\begin{figure}[hbt!]
    \centering
    \includegraphics[scale=0.25]{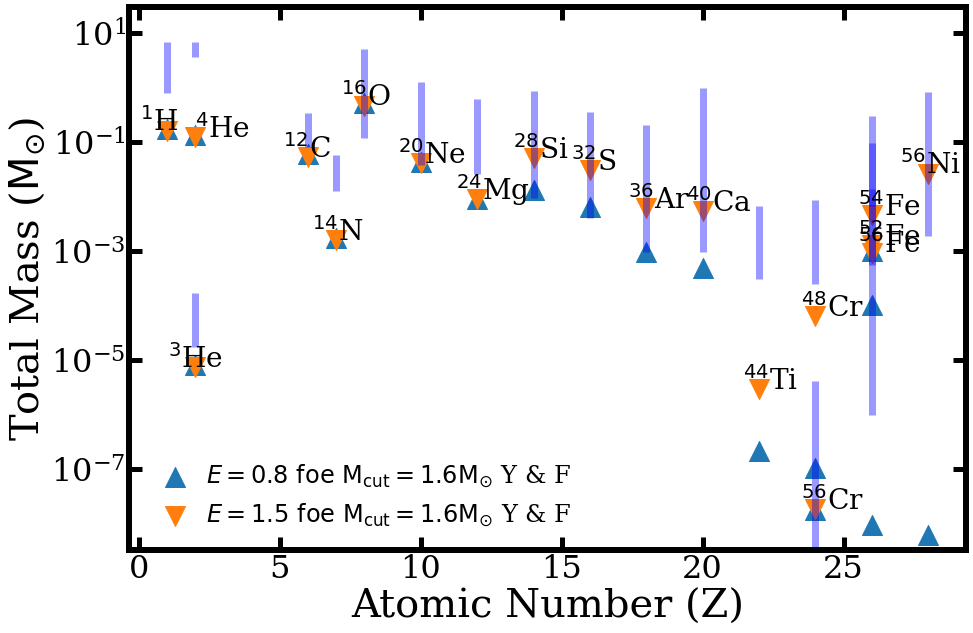}
    \caption{Supernova Elemental Abundances plotted against Results from \citet{YF07}. The blue lines correspond to the yields for the SNe grid. Our abundances differ from YF07 in part because of our much smaller reaction network, and lack of reaction chains containing neutron rich elements.}
    \label{fig:SNeabund}
\end{figure}

\subsubsection{\texttt{snec} Explosion Calibration}
YF07 use a 23$M_{\odot}$ ZAMS mass star as their baseline model, and they also implement a much larger burning network than our models which allows for many more element creation and destruction pathways. The practical impact of using a larger network is in how much of each element is burned in a particular $T-\rho$ regime. For instance, in the \texttt{approx21} network, the photodisintegration reaction $^{20}$Ne($\gamma$, $\alpha$)$^{16}$O reaction \citep{arnett74} enhances $^{24}$Mg by a factor of $\lesssim$ 10 over networks with larger numbers of isotopes. The reason for this excess is that in the more complex networks, neutrino cooling in the burning layers leads to increased densities which can prevent vigorous Ne ignition \citep{farmer16}. 

Similar enhancements exist for $^{28}$Si burning, which burns to form Fe-group elements, including $^{56}$Ni. The \texttt{approx21} network includes a "neutron-eating" isotope of chromium. Owing to a large Coulomb barrier, $^{28}$Si does not generally burn directly to $^{56}$Ni. Instead, the reaction proceeds as first $^{28}$Si photo-disintegration and then a chain of reactions in quasistatical equilibrium (QSE), resulting in Fe-group elements \citep{hix96}. QSE reactions lead to a high degree of neutron-rich isotopes. However in the reduced reaction network we use here, the smaller number of isotopes used leads to a smoother composition profile, as seen in Figure~\ref{fig:final_profile}. In practical terms, this can impact the synthesized X-ray spectrum of the final remnant.

\begin{figure}[hbt!]
    \centering
    \includegraphics[scale=0.18]{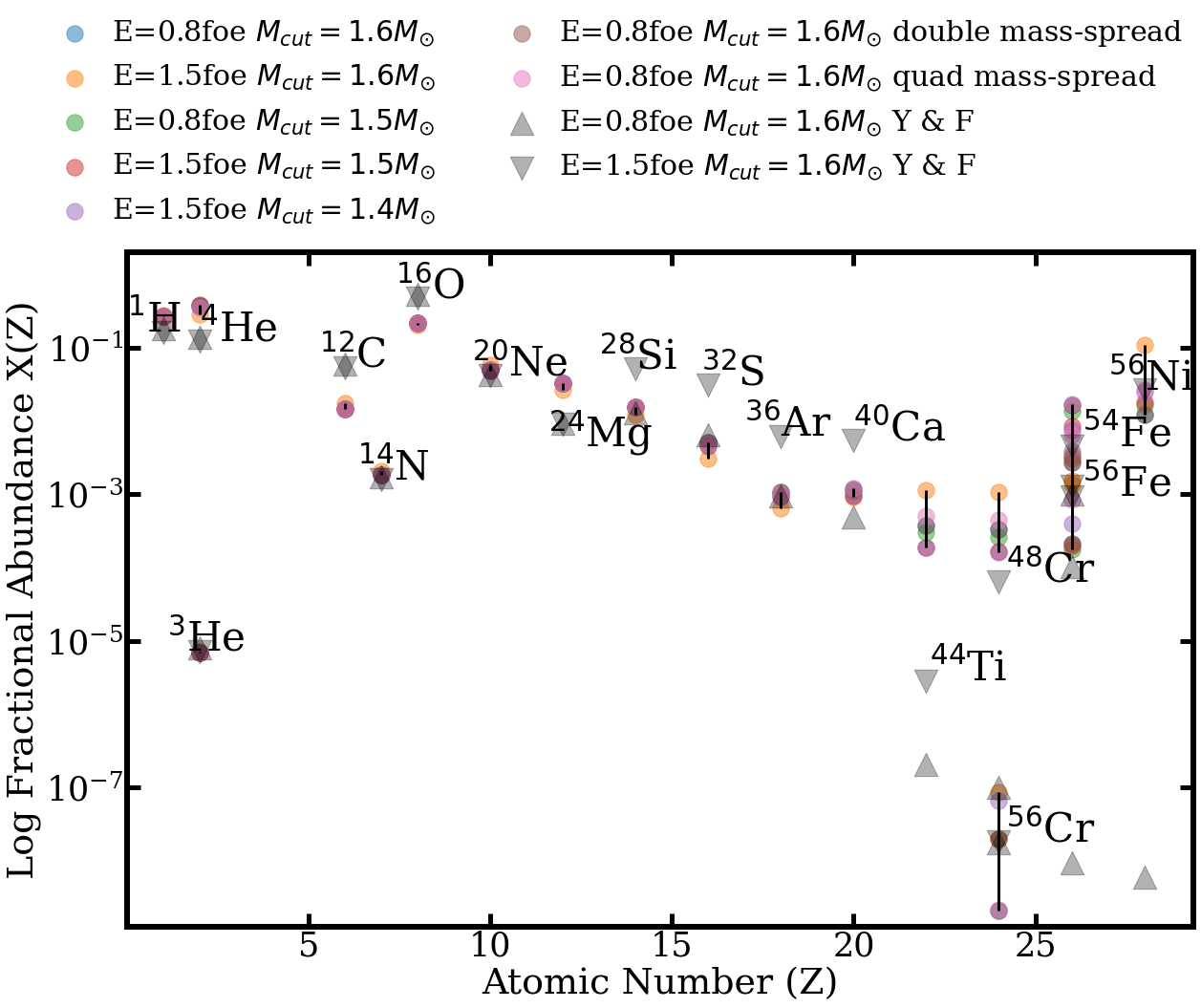}
    \caption{Supernova Elemental Abundances plotted against Results from \citet{YF07} for a $23 \rm M_{\odot}$ exploded using various snec input paramters. Varying the explosion parameters had a more pronounced effect on higher-Z elements.}
    \label{fig:23abund}
\end{figure}

The chief difficulty in calibration was to match the Ni yields following the end of nuclear burning in the SNe. We used the explosion parameters in YF07 as a starting point, and varied the injection time and injection region as well as the delay to align our final Ni yield with that in the literature. We present the yields of the surface rotating models plotted against the yields from YF07 in Figure~\ref{fig:SNeabund}, and present the $23\rm M_{\odot}$ calibration models in Figure~\ref{fig:23abund}. Most elements align reasonably well with our models, although the abundances for some models are significantly higher than expected.  ${}^{24} \rm Mg$ overshoots in part because the reaction network, \texttt{approx21} produces more Mg through longer Ne burning \citep{farmer16}. Additionally, as mentioned above, some isotopes such as ${}^{14} \rm N$ are burned in a limited number of reactions compared to larger nets \citep{approx21,paxton1}.

\subsection{Modeling Supernova Remnants with CSM interactions using \texttt{ChN}}
\label{chn}
\indent Once the supernovae have been simulated out beyond the point of Ni+Co decay ($\sim 100$ days), the resulting profile is piped into the Cosmic-Ray Hydrodynamics Code (\texttt{ChN}). \texttt{ChN} is a one dimensional Lagrangian hydrodynamics code developed out of \texttt{VH1} \citep{vh1}. The code simulates the dynamics of the remnant with initial conditions from the final composition and energies supplied by snec, and self-consistently determines the ionization fractions of shocked material, as well as the thermal and non-thermal spectrum, using self-consistent non-equilibrium ionization calculations \citep{ChN1,ChN2,ChN3}. \texttt{ChN} output is coupled with \texttt{atomdb} and used to generate the broadband line emission from the remnant \citep{hehadb,atomdbconf}. \texttt{ChN} additionally has the ability to simulate the effects of cosmic-ray acceleration on the remnant dynamics, but we defer an conversation those by operating in the so-called test particle limit \citep{ellison07}. We performed simulations to $t_{\mathrm{SNR}}$ = 7000 yrs. Similar work has already been performed for type Ia remnants, and the results have shown ambient density driven differences in the evolution of the remnants \citep{chnIa}.

\begin{figure}[htb!]
    \centering
    \includegraphics[scale=0.24]{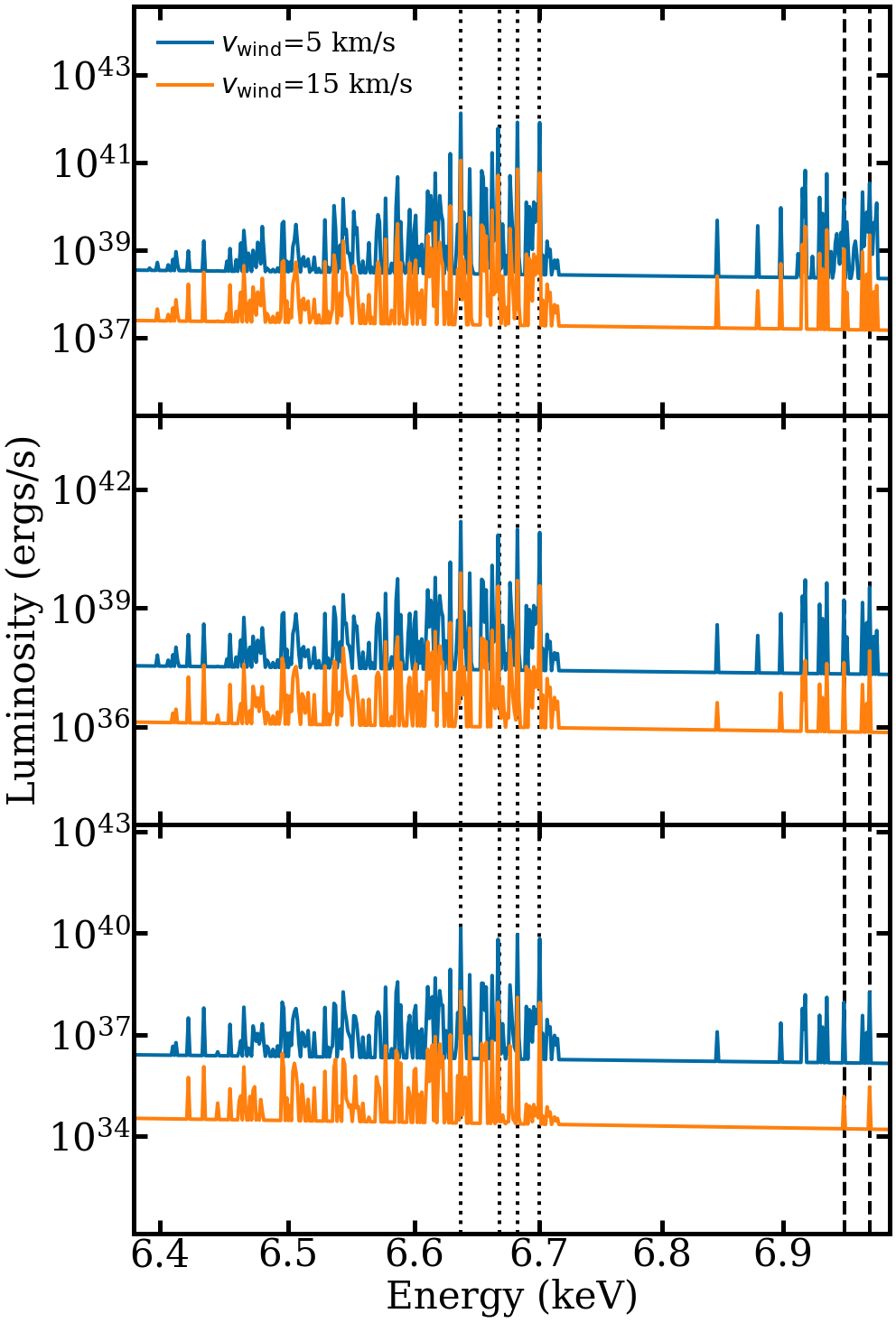}
    \caption{Example Forward Shock (FS) spectra for $\rm M_{\rm ZAMS} = (17.5,~20.6,~28.9)~ \rm M_{\odot}$ at 1000 yrs post-CC (bottom to top). Higher energy emission is particularly dependent on the wind density, with the He-like Fe (dotted) peaks growing significantly as the density increases. The complex between 6.4 keV and He-like Fe is composed of emission from lower Fe ionization states $(\sim \rm Fe~XVII$  upwards) and their satellite lines.  H-like Fe (Ly $\alpha$, dashed) also grows and develops satellite lines with increased density.}
    \label{fig:example-spec}
\end{figure}

Initial plans also included modeling the dynamics of the CSM using \texttt{VH1} and using that final profile as the initial CSM in \texttt{ChN}. In practice, these simulations resulted in main-sequence bubbles several parsecs from the star and what was left within range of the SN shock was not significantly different from taking the average mass-loss rate over the last 500 kiloyears before core-collapse. This is not necessarily true of the stars that exhibited blue loops and had extended periods of lower mass loss, but the effects caused by those changes are not currently considered. Wind velocities from \texttt{mesa} seemed to be systematically under-reported due to issues defining the exact boundary of the photosphere. To combat this, we instead performed two runs for each model with $v_{\rm wind} = (5,15)\  \text{km\ s}^{-1}$. Example spectra for the three models from Figure~\ref{fig:final_profile} are presented in Figure~\ref{fig:example-spec}.

\section{Analysis}
\label{spectra}
\begin{figure}[hbt!]
    \centering
    \includegraphics[scale=0.17]{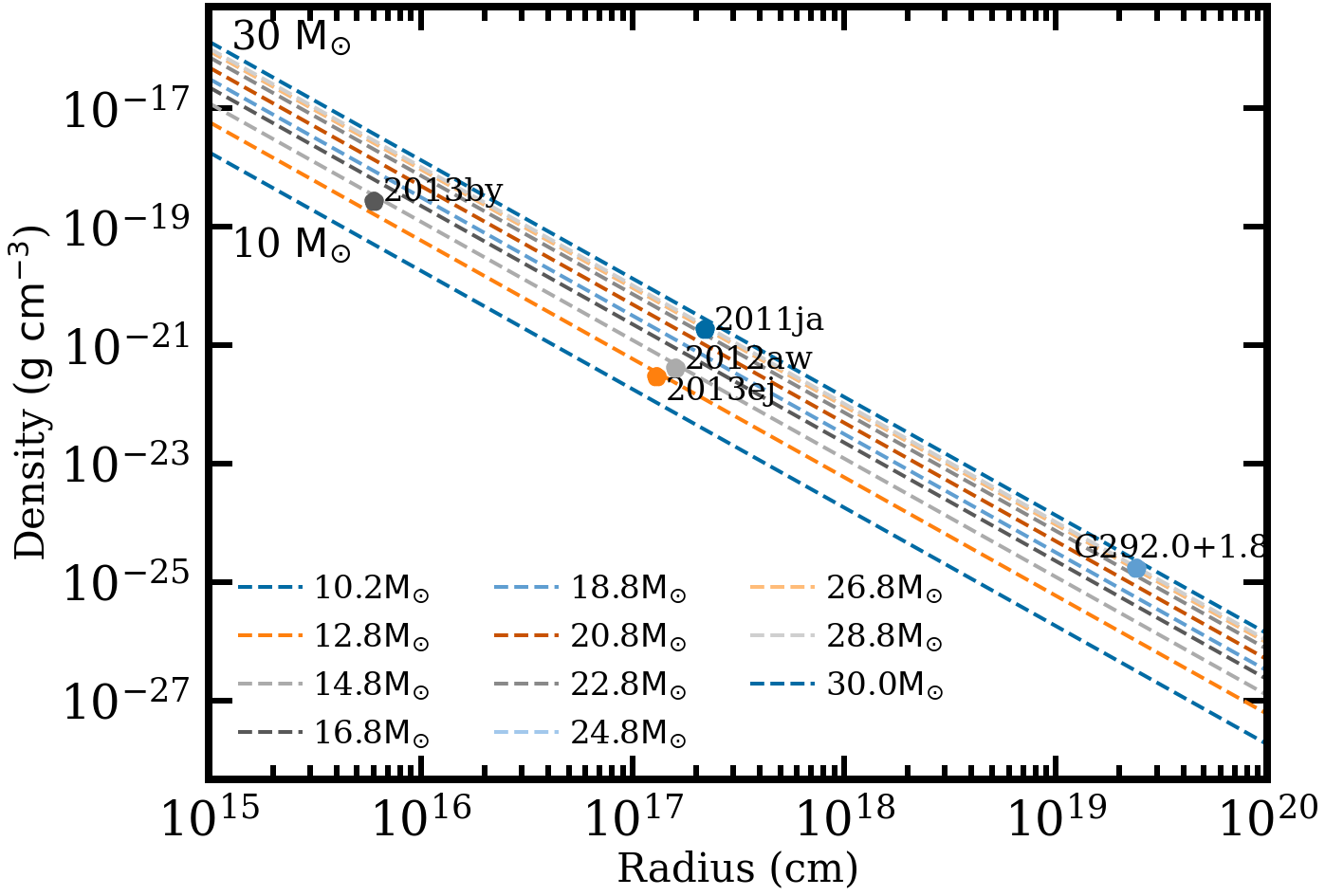}
    \caption{CSM Density profiles for the SNR at Core-Collapse. The wind velocity is taken to be $\rm v_{\rm wind}=~10~\rm kms^{-1}$, making the density dependent on only the model mass-loss rate. Derived CSM densities for various remnants assuming the same RSG wind velocity. The data are in reasonably good agreement with the model densities. SNe data are taken from \citet{chakraborti13,chakraborti16} for SN~2011ja and SN~2013ej, \citet{black17} for SN~2013by, and \citet{yadav14} for SN~2012aw. The inferred mass loss rate for G292.0+1.8 is taken from \citet{jjlee10}.}
    \label{fig:wind}
\end{figure}

In spite of the fact that the models shown in Figures~\ref{fig:final_profile} and  \ref{fig:example-spec} have approximately the same final mass, their spectra are noticeably different. One reason for this is that 
most SNe and SNR are found to evolve into a range of density profiles which are dictated by the progenitor mass-loss rate and the wind velocity, and the X-ray emission is dependent upon the circumstellar density. This is shown in Figure~\ref{fig:wind} where we use the inferred mass-loss rates from our models to construct the circumstellar profile around the progenitor at the time of core-collapse, assuming a wind velocity of $10~\rm km s^{-1}$. We also include inferred mass loss rates for a selection of SN~IIP/IIL and the Galactic SNR G292.0+1.8. 

As seen in Figure~\ref{fig:wind}, increasing progenitor mass, in broad terms, maps to increased mass loss rates during the later stages of evolution. Mass-loss rates in SN~IIP/IIL and some Galactic SNRs are inferred from radio and X-ray observations, and are in broad agreement with the rates used here. For the larger rates discussed in \citet{forster}, The enhanced mass loss would lie within $~10^{14} \rm cm$, and would have a larger affect during the very early stages of SN evolution, whereas the RSG winds drive the SN and SNR evolution at later epochs. Additionally, the red supergiant wind will interact with the low density wind from the main sequence star. The RSG wind sweeps the main sequence wind material into a radiatively cooled shell located $\sim$ 10$^{20}$ cm from the progenitor \citep{castor75}. Models which extend into the boundary region between the red supergiant and main sequence winds will also need to consider radiative losses, as the swept up shell of material can rapidly decelerate the blastwave. This is an important consideration for remnants with ages $\gtrsim$ 10$^{4}$ yrs, but we do not consider it here.

\subsection{Calculating Centroid Energies and Ionization States}
\label{ionization}
\begin{figure}[htb!]
    \centering
    \includegraphics[scale=0.19]{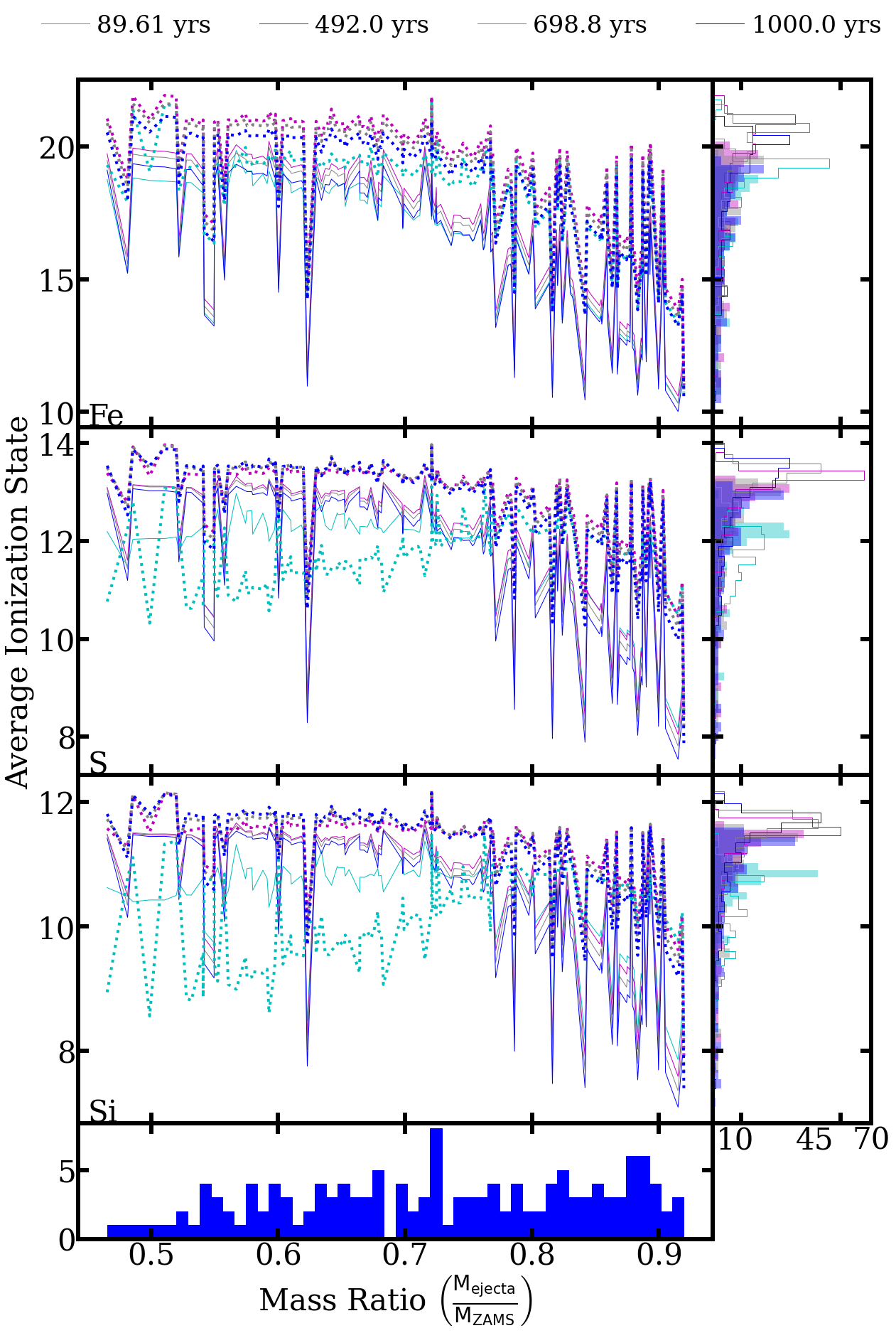}
    \caption{Average ionization states in the shocked region for the elements Fe, S, and Si (top to bottom) in a fast wind $(\rm solid,~v_{\rm wind}=~15~\rm kms^{-1})$, and a slow wind $(\rm dashed,~v_{\rm wind}=~5~\rm kms^{-1})$. The infilled histograms correspond to the charge states of the fast wind, and the outlined histograms correspond to the slow wind. The bottom histogram shows the distribution of mass ratios for the remnant grid. Lower Z elements are quickly pushed to high ionization states and exhibit a longer recombination time. Ionization in the slow wind models appears to occur more slowly than ionization in the fast wind, but ultimately reaches a higher charge state. This is due to the higher density material leading to more collisional ionization.}
    \label{fig:fast-ion-combo}
\end{figure}

X-ray spectra in supernova remnants encode physics about the chemical composition of the ejecta and the evolution of the progenitor. Line emission is produced by ionization and excitation of shocked circumstellar and ejecta material.  In the case of SNRs, Ionization and recombination rates are driven by Coulomb interaction, which are directly proportional to the CSM density.  Photoionization is not currently considered. The ionization state of each element is also driven by electron number, with higher-Z elements requiring more energy to fully ionize, while more readily recombining as can be seen in Figure~\ref{fig:fast-ion-combo}. Therefore, line emission from these highly-ionized high Z elements strongly tracks the evolution of the remnant on timescales suitable for probing the late phases of stellar evolution. 

In order to better examine the trends that result from the CSM, Figure~\ref{fig:fast-ion-combo} and later figures that connect ionization to progenitor mass are plotted against the mass ratio
\begin{equation}
    \rm Mass\ Ratio = \frac{\rm M_{\rm final\ model}- M_{\rm excised}}{M_{ZAMS}}
\end{equation}
Unlike the ZAMS mass, the mass ratio is sensitive to the mass-loss rate of the individual stars, so stars with relatively low mass-loss rates (ie. the blue loops) and low CSM densities will be grouped together. Looking at the ratio also allows for the larger final progenitor mass to be included in these analyses as the mass-loss rates for many blue loop stars are comparable to lower mass objects but the ratio will be higher due to due to the much larger ejecta mass in the blue loop case. The differences in remnant composition do have implications for the ionization and spectral behavior, so it is important to look at metrics that consider all of these features.

\begin{figure}[htb]
    \centering
    \includegraphics[scale=0.18]{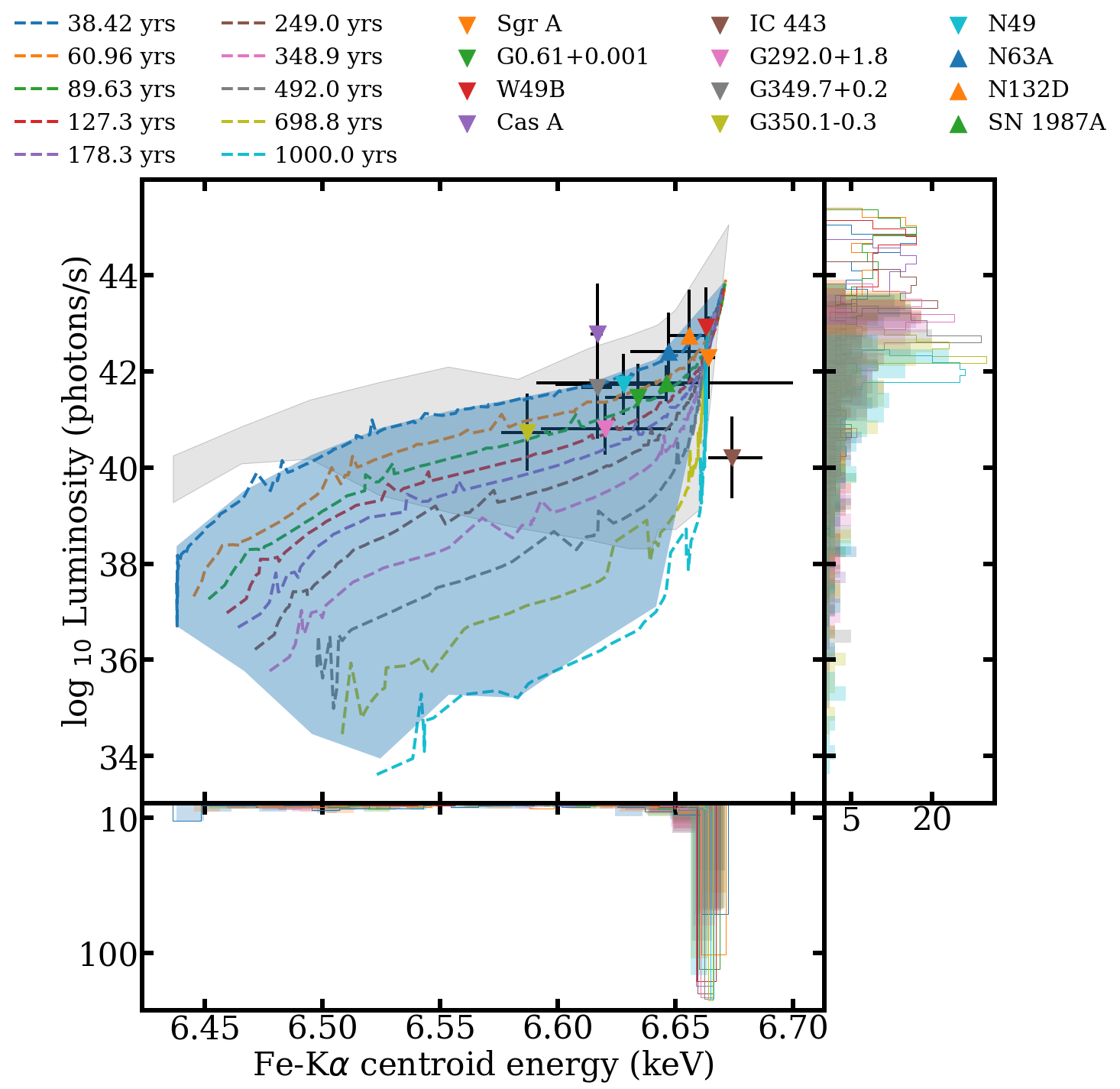}
    \caption{Fe-K$\alpha$ centroids for the fast wind ($v_{\rm wind}=~15~\rm kms^{-1}$) models. The centroid luminosities are plotted against centroid energy for the: forward shock (blue surface), the reverse shock (grey surface), and model isochrones for the forward shock (dashed lines). FS values correspond to the infilled histograms, and RS values to the outlined histograms.}
    \label{fig:fast-cent-fs}
\end{figure}

\begin{figure}[htb]
    \centering
    \includegraphics[scale=0.18]{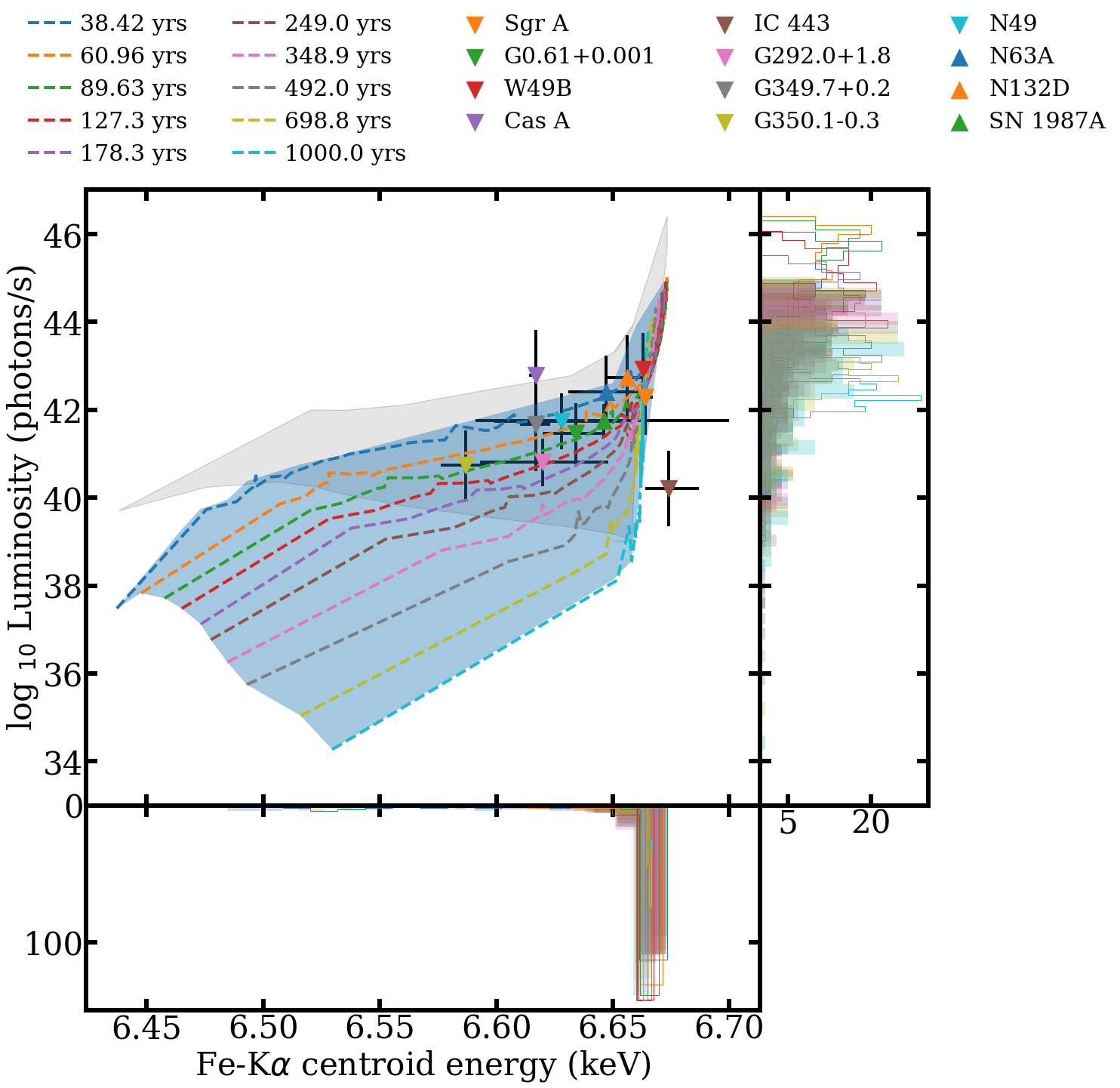}
    \caption{Fe-K$\alpha$ centroids for the slow wind ($v_{\rm wind}=~5~\rm kms^{-1}$) models. The centroid luminosities are plotted against centroid energy for the: forward shock (blue surface), the reverse shock (grey surface), and model isochrones for the forward shock (dashed lines). FS values correspond to the infilled histograms, and RS values to the outlined histograms.}
    \label{fig:slow-cent-fs}
\end{figure}

Higher ionization states of Fe have proven to be useful probes of remnant evolution due to the relatively high elemental abundance in the remnant and the relative isolation of the K-shell emission in the X-ray spectrum.  Fe-K$\alpha$ centroids have also been cited as a probable discriminant of SNR progenitor type \citep{yamaguchi}. Core-collapse remnants show a strong trend toward the He-like transitions, while Type Ia centroids are dominated by Fe XVII and other similar ionization states. Our models support the general trend of core-collapse remnants exhibiting harder Fe-K$\alpha$ centroid energies, with models trending toward the Fe XXV resonance transition as the remnant ages. We do note that the circumstellar density is the strongest determinant of an emission line's centroid energy and luminosity, as can be seen in Figures~\ref{fig:fast-cent-fs} and~\ref{fig:slow-cent-fs}. 
Because of this strong density dependence, lower mass models can exhibit softer centroid energies, especially when the remnant is still young. These centroids do still harden in time, and after $\sim$ 1000 years they approach the energy which distinguishes SNR Ia from CC SNR, as defined in \citet{yamaguchi}.

\begin{figure}[htb]
    \centering
    \includegraphics[scale=0.21]{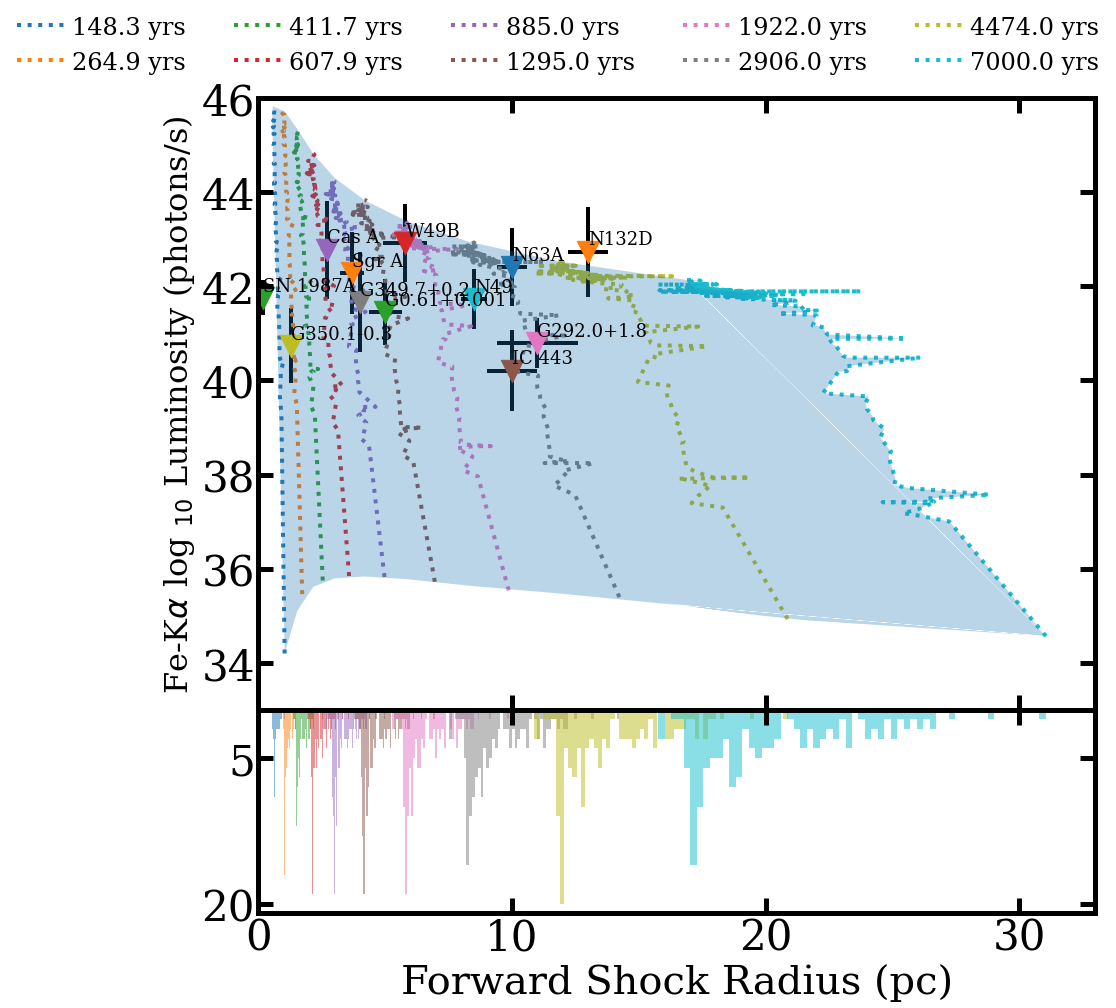}
    \caption{Fe-K$\alpha$ centroid luminosities for the fast wind ($v_{\rm wind}=~15~\rm kms^{-1}$) models. The luminosities are plotted against remnant size for the total shocked region (blue surface) along with model isochrones for the shocked region (dashed lines).}
    \label{fig:slow-rad-fs}
\end{figure}

\begin{figure}[htb]
    \centering
    \includegraphics[scale=0.21]{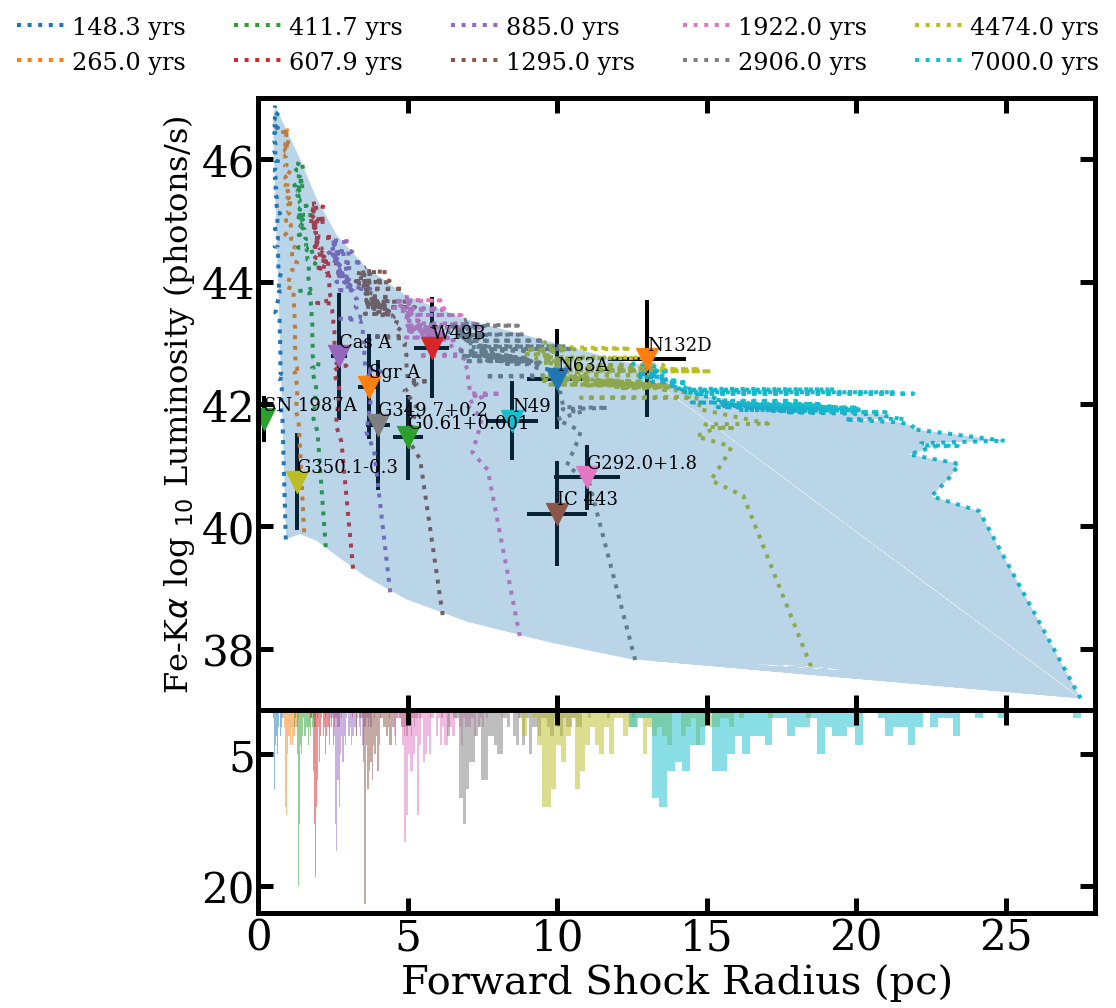}
    \caption{Fe-K$\alpha$ centroid luminosities for the slow wind ($v_{\rm wind}=~5~\rm kms^{-1}$) models. The luminosities are plotted against remnant size for the total shocked region (blue surface) along with model isochrones for the shocked region (dashed lines).}
    \label{fig:fast-rad-fs}
\end{figure}

Examining remnant luminosity as a function of remnant size yields additional insight into the mass-loss history of SNR. As can be seen in Figure~\ref{fig:slow-rad-fs}, and Figure~\ref{fig:fast-rad-fs}, the Fe-K$\alpha$ luminosity also correlates well with FS radius for a given remnant age. Comparing these results to spectral data yield remnants which seem to be reasonably well constrained by a wind-driven model. As an example, the centroid energy, luminosity, and remnant size for IC 443 seem to agree with the derived age of $\sim 3000$ years and a ZAMS mass of $\sim 16-25 \rm M_{\odot}$ from  \citet{troja08}. 

Even in cases where remnants do not follow the expected behavior of our models, we can still infer some information about the mass-loss history. For instance, in the case of objects where the centroid energy is softer than expected, given its measured size and inferred age, we can infer that the Fe in the remnant is not highly ionized, which can be interpreted as meaning a lower density in the shocked region than is indicated by the wind model. Lower than expected Fe-K$\alpha$ luminosity at a given centroid energy can also indicate a lower than modeled Fe abundance in the shocked region.

The current generation of X-ray observatories are capable of measuring the Fe-K$\alpha$ centroid because of its luminosity and relative isolation from other lines. Next generation observatories, such as \textit{XRISM} and \textit{Athena}, will be able to resolve lines in more crowded regions of the X-ray spectrum. Many lower-Z elements such as Mg or Si still have issues related to either being ionized beyond the He-like state, or with other emission lines overlaying the transitions, but one possible element that is sufficiently separated, while maintaining a significant population of He-like ions is S, as discussed in \S~\ref{sec:snr_obs}.

\subsection{Hydrogen-like to Helium-like Line Ratios}
\label{hhefs}
\begin{figure}[htb]
    \centering
    \includegraphics[scale=0.17]{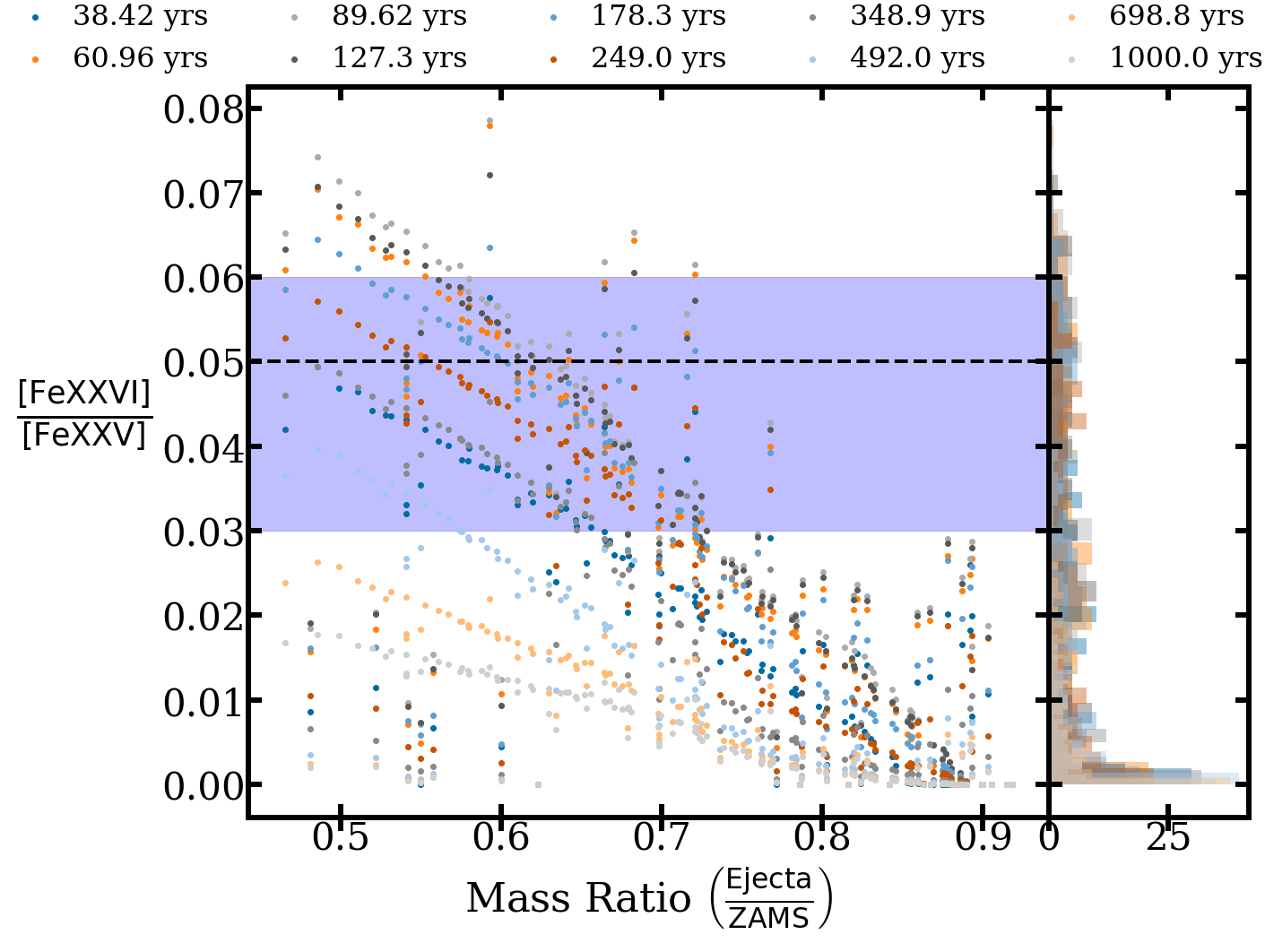}
    \caption{H-like to He-like line ratios for the fast wind case. The horizontal dashed line and shaded region correspond to a measurement of the line ratio for Sgr-A East \citep{koyama}. The overall trend is driven by the much shorter recombination time and much higher ionization temperature of Fe XXVI compared to Fe XXV.}
    \label{fig:fast-hhe-fs}
\end{figure}

\begin{figure}[htb]
    \centering
    \includegraphics[scale=0.17]{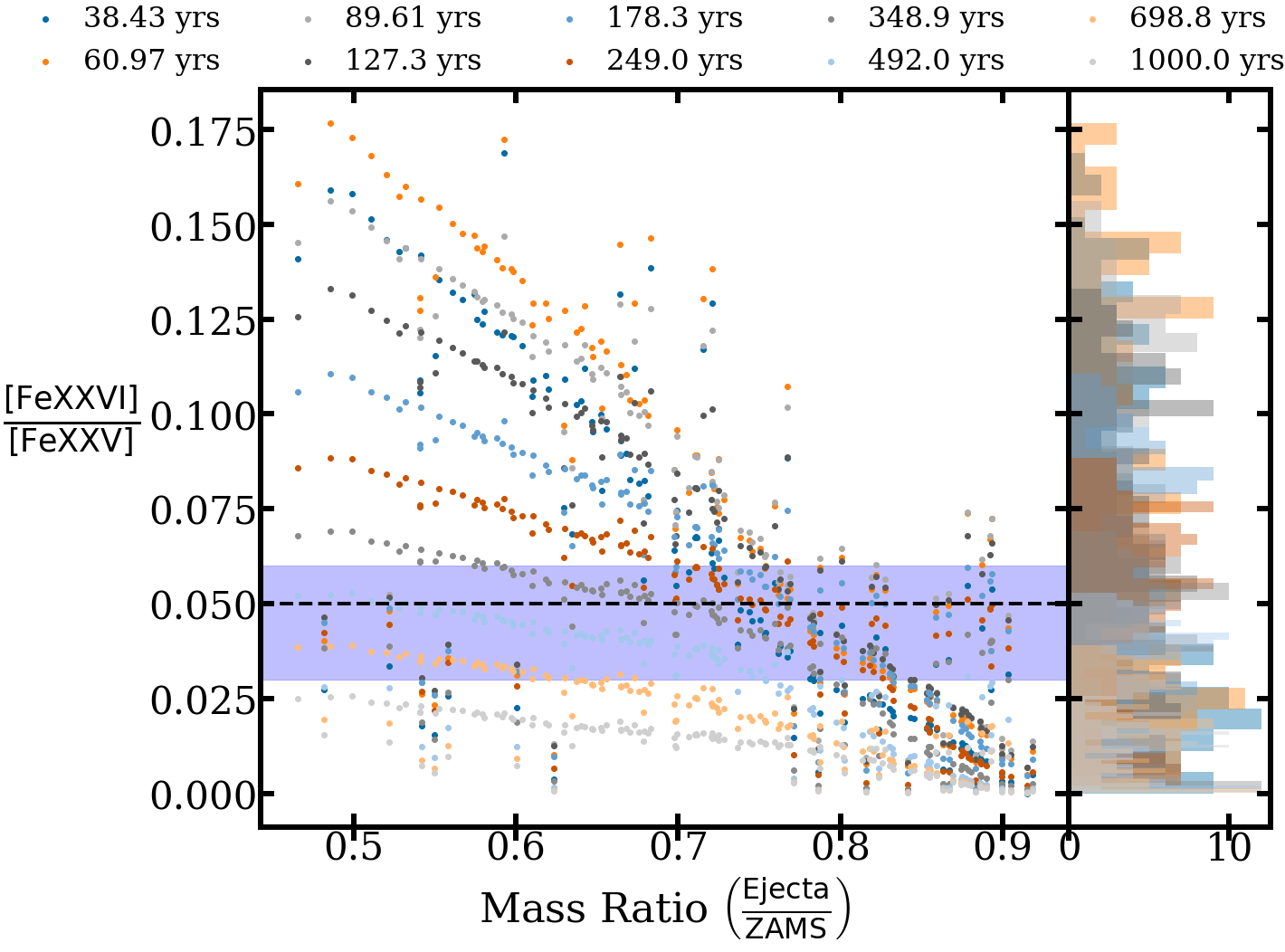}
    \caption{H-like to He-like line ratios for the slow wind case. The horizontal dashed line and shaded region correspond to a measurement of the line ratio for Sgr-A East \citep{koyama}. The mechanism and trend are similar to the one described in Figure~\ref{fig:fast-hhe-fs}, but the increased CSM density leads to increased formation of Fe XXVI compared to the fast wind case.}
    \label{fig:slow-hhe-fs}
\end{figure}

Separating the Fe-K$\alpha$ and Fe-Ly$\alpha$ lines is something that is within the ability of some current X-ray missions. Comparing the relative line strengths of Fe XXV (Fe-K$\alpha$) and Fe XXVI (Fe-Ly$\alpha$, Fe-Ly$\beta$) emission lines can be used to probe the ionization state of the remnant. In the case of models exhibiting wind-driven mass loss, the H-- to He-like line ratios also show a dependence on the ZAMS to ejecta mass ratio of the model. Higher mass ratio models correspond to lower mass-loss rates, and lower circumstellar densities. We present the ratios of the He-- to H-like lines in the shocked circumstellar material in Figures~\ref{fig:fast-hhe-fs}, and \ref{fig:slow-hhe-fs}. The line ratios in general decrease as a function of ZAMS to ejecta mass ratio and also as a function of time. 
 
 In Figures~\ref{fig:fast-hhe-fs} and~\ref{fig:slow-hhe-fs}, we also include a measured line ratio from \citet{koyama} for Sgr-A East. The measurement is in the range of our models, although it seems to imply a younger than expected remnant. The H-like to He-like ratio puts it at $\sim 400-900~\rm yrs$, while the Sgr-A East centroid and radius values put it at around $\sim 1000 \rm~yrs$ old, which is consistent with the literature \citep{koyama}. Sgr-A East seems to be over-ionized compared to our models, which also seems to agree with the literature. We do note though that it is possible Sgr-A East is not a CC SNR, but instead the product of a Type Iax SN, which might explain the high line ratio \citep{zhou20}. 

\section{Comparisons with SNR Properties}
\label{sec:snr_obs}
While the Fe-K line centroid and luminosity of SNRs is a good discriminator between
core collapse and Type Ia remnants, not all core collapse supernova remnants have
strong Fe-K line emission. In this section, we explore how some of the other 
SNR spectral and dynamical properties compare with our models. We focus on a handful
of relatively young objects which have been previously studied in detail with 
{\it Chandra}. We explore how the He-like sulfur line centroid can be used to 
discriminate between models, as it is typically bright in SNR X-ray spectra, even
when iron is not.

We list the handful of objects we explore in Table~\ref{tab:snrprop}. For
each SNR, we fit the sulfur He-like triplet to a Gaussian, the normalization of which
is simply the line flux. In order to convert the line flux to a luminosity, distances to each object are pulled from the literature. 

\begin{deluxetable*}{lcccccr}
\tablecolumns{8}
\tablewidth{0pc}
\tablecaption{Properties of Sulfur He-like triplet for Selected Supernova Remnants 
\label{tab:snrprop}}
\tablehead{
\colhead{SNR} & \colhead{Distance} & \colhead{R$_{\mathrm{FS}}$} & \colhead{t$_{\mathrm{SNR}}$\tablenotemark{a}} & \colhead{L$_{\mathrm{X}}$\tablenotemark{b}} & \colhead{Centroid Energy} & \colhead{Ref.} \\
\colhead{} & \colhead{kpc} & \colhead{pc} & \colhead{yr} & \colhead{10$^{42}$ photons s$^{-1}$} & 
\colhead{keV} & \colhead{}
}
\startdata
1E~0102-7219 & 62 & 6.3 $\pm$ 0.2 & 1738 $\pm$ 175 & 17.8 $\pm$ 3.6 & 2.445 $\pm$ 0.009 & 1,2 \\
N132D & 50 & 12.5 $\pm$ 0.2 & 2450 $\pm$ 195 & 173.8 $\pm$ 15. & 2.439 $\pm$ 0.002 & 3,4 \\
Cas A & 3.4 $\pm$ 0.3 & 2.6 $\pm$ 0.1 & 340 $\pm$ 15 & 7.76 $\pm$ 0.4 & 2.449 $\pm$ 0.001 & 5 \\
G11.2-0.3 & 5.0 $\pm$ 0.5 & 2.6 $\pm$ 0.3 & 1400 -- 2400 & 2.88 $\pm$ 0.5 & 2.430 $\pm$ 0.002 & 6 \\
Kes 79 & 7.1 $\pm$ 0.6 & 12.4 $\pm$ 0.8 & $\sim$ 6000 & 0.2 $\pm$ 0.1 & 2.435 $\pm$ 0.007 & 7 \\
Kes 73 & 8.5 $\pm$ 0.8 & 6.3 $\pm$ 0.3 & $\sim$ 1800 & 4.26 $\pm$ 0.5 & 2.435 $\pm$ 0.002 & 8 \\
Kes 75 & 19 $\pm$ 2.0 & 9.7 $\pm$ 1.0 & $\sim$ 1000 &  4.30 $\pm$ 0.75 & 2.438 $\pm$ 0.003 & 9 \\
RCW 103 & 3.8 $\pm$ 0.4 & 5.0 $\pm$ 0.5 & $\sim$ 4400 & 0.81 $\pm$ 0.1 & 2.431 $\pm$ 0.002 & 10 \\
G292.0+1.8 & 6 $\pm$ 0.5 & 7.7 $\pm$ 0.8 & 3000 -- 3400 & 8.2 $\pm$ 0.1 & 2.442 $\pm$ 0.003 & 11,12\\
\enddata
\tablenotetext{a}{SNR ages are determined either by optical or X-ray proper motion studies of ejecta, 
or with comparisons to self-similar models. Those with ranges or which are given as approximate values were
estimated via model comparisons.}
\tablenotetext{b}{X-ray luminosity is for the 2.35 -- 2.55 keV range.}
\tablerefs{[1]: \citet{xi19}; [2]: \citet{banovetz21}; [3]: \citet{sharda20}; [4]: \citet{law20}; 
[5]: \citet{pat15}; [6]: \citet{borkowski16}; [7]: \citet{sun04}; [8]: \citet{borkowski17}; 
[9]: \citet{chevalier05}; [10]: \citet{braun19}; [11]: \citet{ghavamian05}; [12]: \citet{bhalerao19}}
\end{deluxetable*}

In total, we examine 2800 model spectra with ages to 7000 years and radii out to $\lesssim$ 30 pc, though we note that models which extend to these larger radii may also interact with swept-up main sequence wind, which we do not consider here. This is a large, but finite set of
modeled data, and given that simulation produces X-ray spectra on a logarithmic
timescale, most of our modeled spectra are skewed to younger ages. We therefore 
use a Gaussian Mixture Model to represent the distribution of our simulated spectra,
and fill in the gaps. This is discussed in detail below. 

\begin{figure*}
    \centering
    \includegraphics[width=0.45\textwidth]{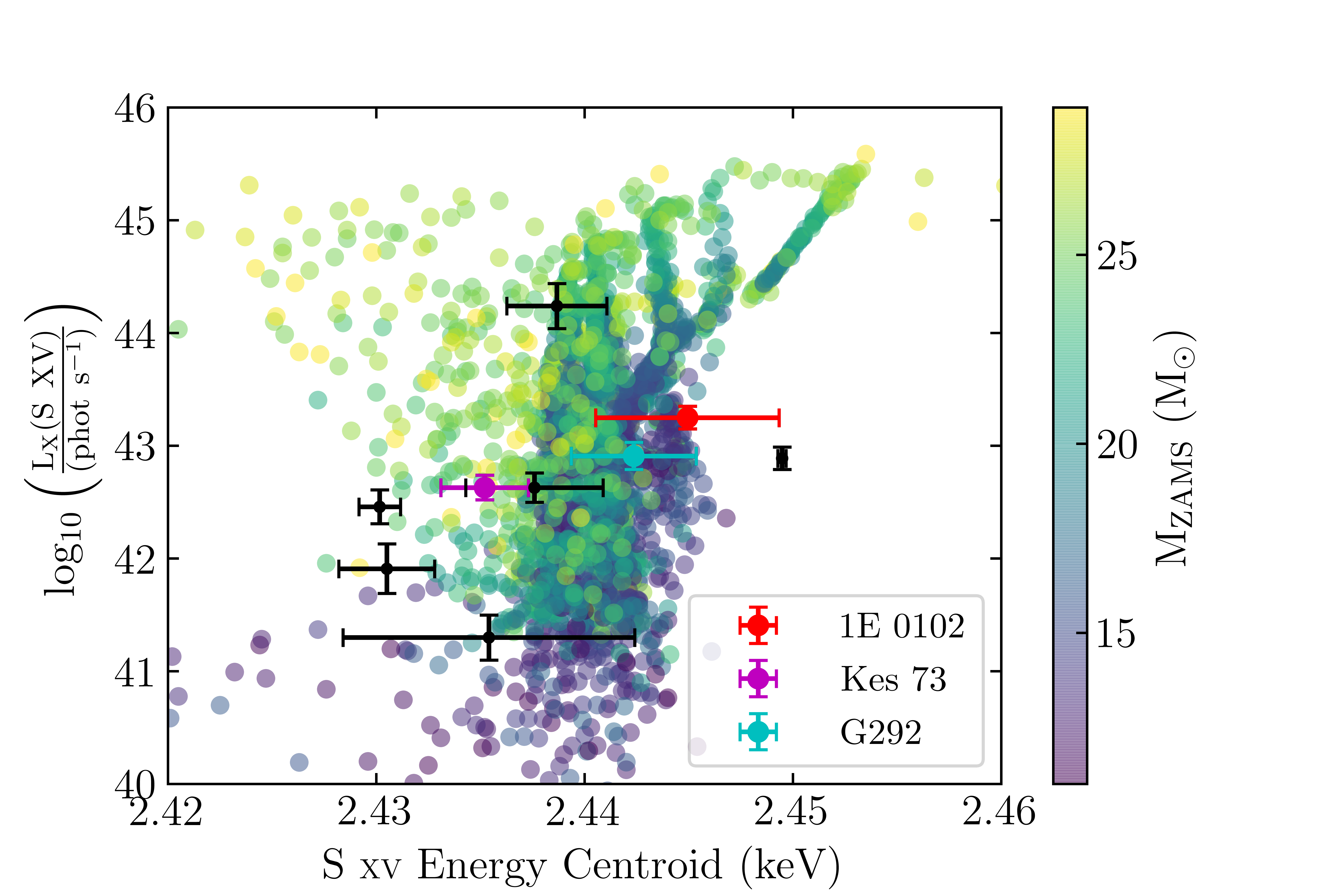}
    \includegraphics[width=0.45\textwidth]{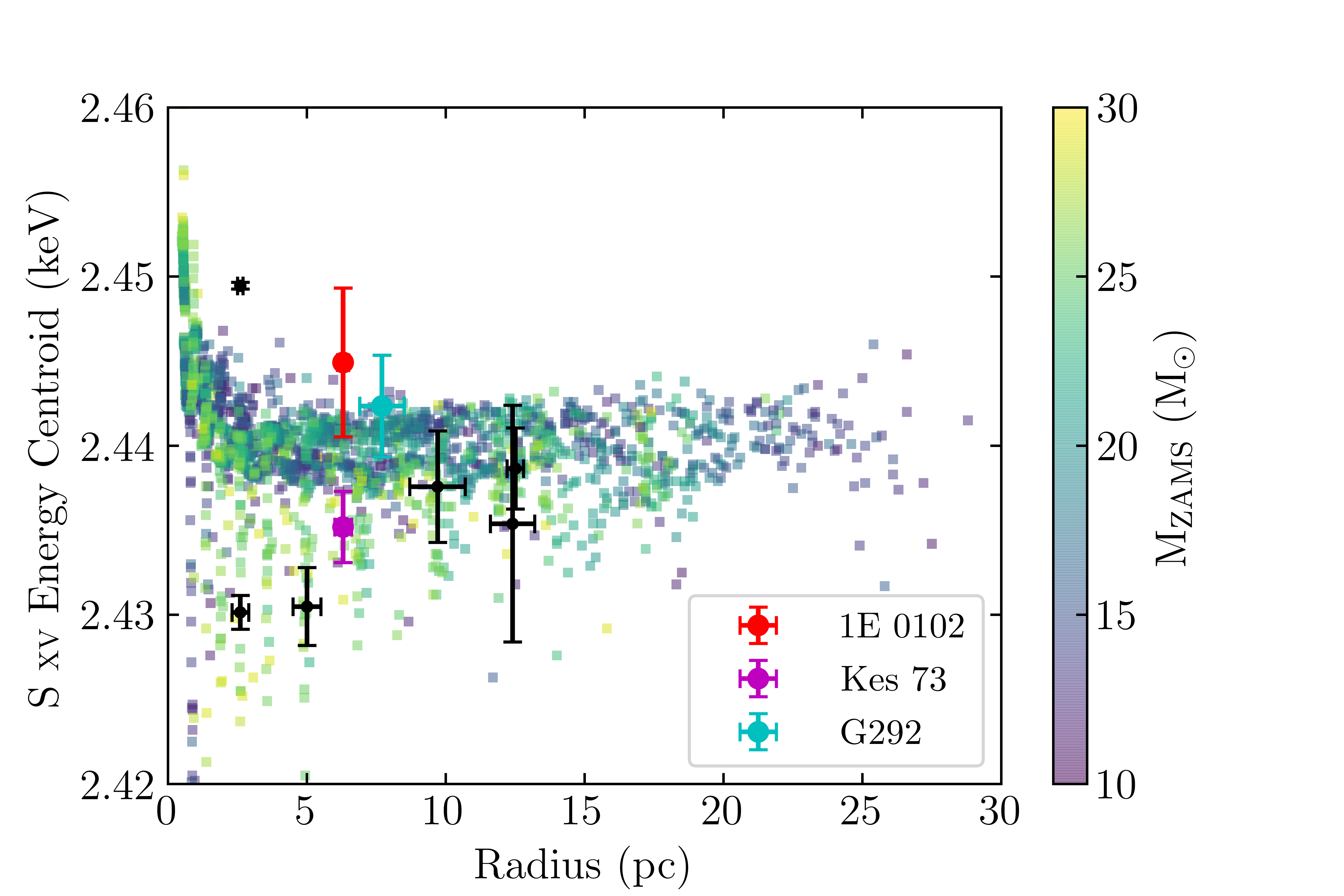}
    \caption{{\bf Left:} He-like sulfur line centroid for our models, colored as a function of progenitor mass. The distribution includes both the fast and slow wind cases, and includes models to ages of 7000 years. {\bf Right:} He-like line centroid vs model forward shock radius. In both cases, the data from Table~\ref{tab:snrprop} are overplotted. We highlight the measured properties for 1E 0102-7219, Kes 73, and G292.0+1.8, which are discussed in the text.
    \label{fig:s_properties}}
\end{figure*}

For each modeled spectra, we use \texttt{XSPEC} to simulate a 
{\it Chandra}
ACIS-S observation of the simulated remnant. We then fit the He-like line in the
simulated spectrum to a Gaussian, in order to get the triplet centroid and luminosity. These data are shown in Figure~\ref{fig:s_properties}, where we plot the line centroids and luminosities as a function of forward shock radius and progenitor ZAMS mass. We also overplot the measured spectral characteristics of SNR listed in Table~\ref{tab:snrprop}.

As seen in Figure~\ref{fig:s_properties}, the simulations show some clustering when comparing the spectral properties to one another (Fig.~\ref{fig:s_properties}, left) or when comparing spectral properties to dynamical characteristics. Additionally, when classifying by ZAMS mass, the lower mass models occupy a distinct part of parameter space, while higher mass models occupy a separate region. For instance, when comparing line centroids to line luminosities, with only a few exceptions the models with M $>$ 25 M$_{\odot}$ tend to lie in the part of parameter space with higher luminosities but lower centroids. Likewise, with a handful of exceptions, those progenitor models with M$_{\mathrm{ZAMS}}$ $\lesssim$ 15 M$_{\odot}$ tend to trend towards lower luminosities, but higher line centroids. This trend seems largely independent of the remnant age. 

In terms of the line centroids vs radius, clustering in the models is a little less clear, but there is a clear spread in line centroid at low radius (Fig.~\ref{fig:s_properties}, right). This clustering of the model data suggests that the distributions of these parameters can be simulated as a mixture of underlying kernels. We can then plot measured variables like those listed in Table~\ref{tab:snrprop}, without being distracted by gaps in the simulated data. To do this, we first assume that the variables we are comparing, the ZAMS mass, model radius, measured sulfur line centroid, and measured sulfur line luminosity, can be represented as smoothly varying functions. The ZAMS mass is our fundamental input variable, and all other quantities derive from it. The assertion that our model grid is drawn from a smooth, underlying kernel is reinforced by the fact that the SNR ejecta evolution and subsequent ionization of material are continually varying funtions of the SNR age. 

We begin by modeling the distributions of variables named above as a mixture of Gaussians. Since we are trying to model the underlying distribution which may not necessarily be Gaussian, the number we choose can be arbitrary. However, we can use the so-called {\it Bayesian information criterion} \citep{liddle07} to compute the likelihood that the underlying distribution is the sum of a number of Gaussians. We iterate on the intial choice of underlying density functions, and then compute the maximum likelihood that that number of models describes the data. In the case of our models, we find that the criterion converges at 13 Gaussians. 
\begin{figure*}
    \centering
    \includegraphics[width=0.45\textwidth]{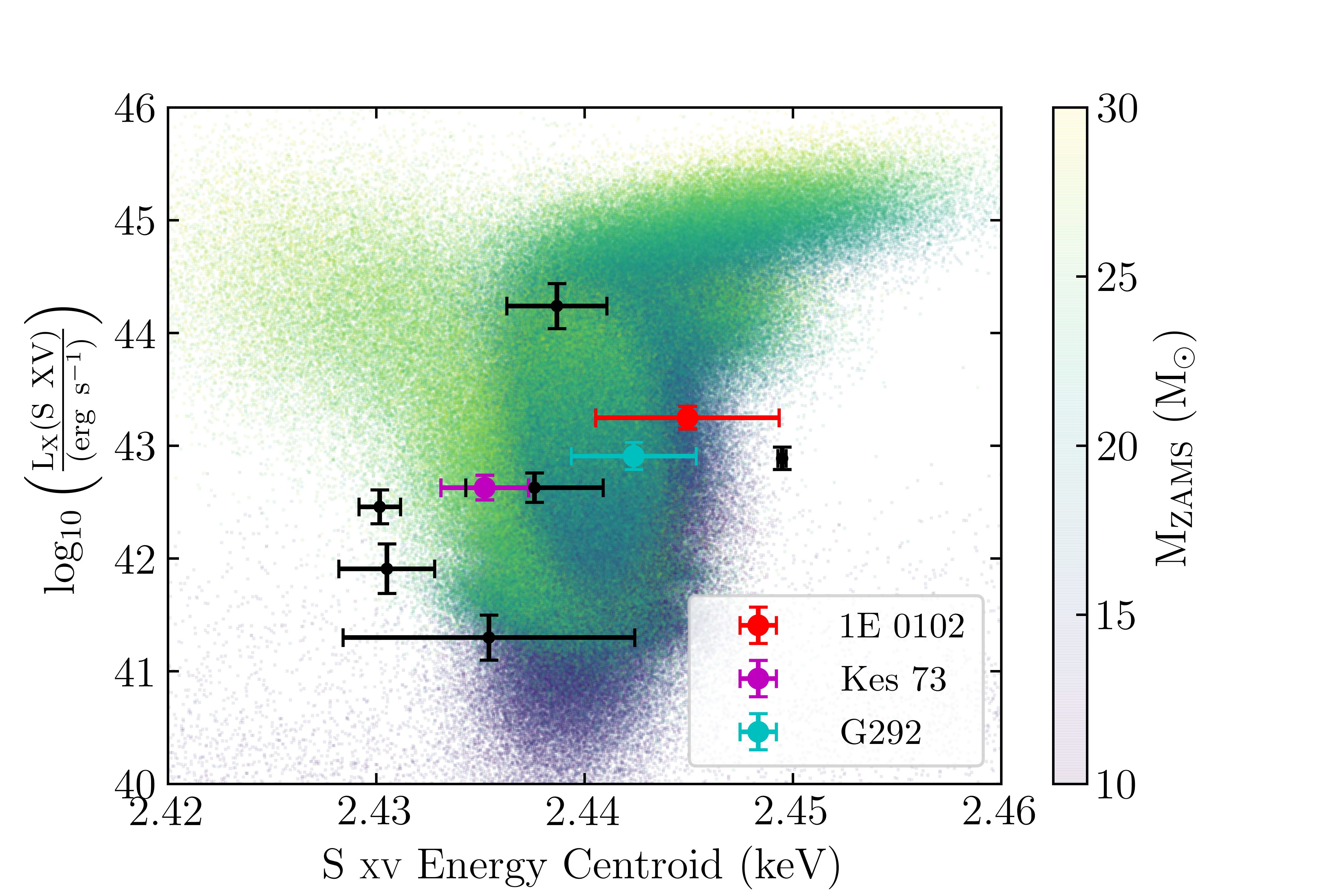}
    \includegraphics[width=0.45\textwidth]{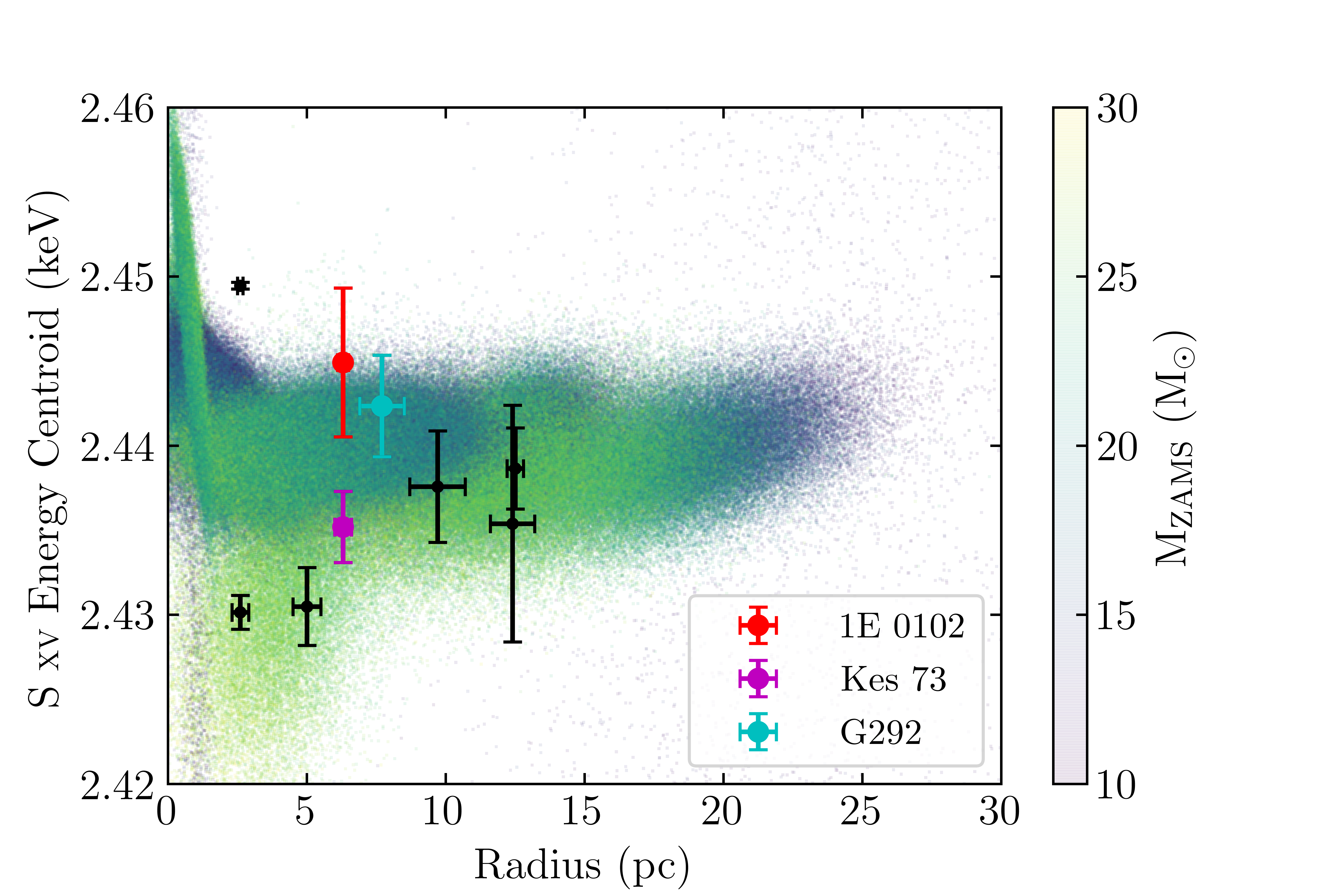}
    \caption{Gaussian Mixture Model representation of the simulated supernova remnant properties, colored by
    ZAMS mass. {\bf Left:} He-like sulfur line centroid versus luminosity. {\it Right:} He-like line centroid
    versus modeled forward shock radius. In both cases, the data from Table~\ref{tab:snrprop} are overplotted. We highlight the measured properties for 1E 0102-7219, Kes 73, and G292.0+1.8, which are discussed in the text.
    \label{fig:s_prop_gmm}}
\end{figure*}
Upon determining the best number of models to use to describe the modeled properties, we replicate the distribution
in order to see where the data are clustered and how these clusters relate to observed SNR properties. 

In Figure~\ref{fig:s_prop_gmm}, we show the same model properties as in Figure~\ref{fig:s_properties}, except now we show the distribution which is cloned from our model distribution. As seen in the left hand side of Fig.~\ref{fig:s_prop_gmm}, there is definite clustering in the model spectral properties. When looking at the cloned 
distribution of forward shock radii vs He-like line centroid, the clustering becomes more apparent. In particular, the higher mass progenitor models tend to have lower line centroids than the lower mass models. It is important to note that this distribution includes both fast and slow winds, and that is revealed in Fig.~\ref{fig:s_prop_gmm} (right), at the smallest radii, where there is a thin strip of high mass models at small radius, overlying 
a broader strip of low mass models, while at low centroid energy, there is a thin strip of low mass models. Conceptually, this suggests that the low mass models which expand into their low density winds do not efficiently heat and ionize the ejecta, and conversely the high mass models in dense winds quickly ionize the ejecta.

We conclude this section by quantitatively comparing our models to the properties of some of the SNR listed in Table~\ref{tab:snrprop}. We begin with the young O-rich SNR 1E 0102-7219, located in the Small Magellanic Cloud. By measuring the proper motion of several undecelerated optical knots, \citet{banovetz21} was able to derive an age of $\sim$ 1700 yrs, while \citet{xi19} estimated a progenitor mass of 15--40~M$_{\sun}$, based on the amount of circumstellar material they estimate the forward shock to have swept up. With a He-like
sulfur line luminosity of 1.8$\times$10$^{43}$ photons s$^{-1}$ and a centroid energy of 2.445 keV, it lies along the boundary between ZAMS models with masses of 15--25 M$_{\odot}$. Comparing its forward shock radius to its sulfur line centroid (Fig.~\ref{fig:s_prop_gmm}, left) also suggests that the progenitor mass was $\sim$ 20 M$_{\sun}$. Additionally, the high line centroid energy suggests that it is interacting with a dense wind. \citet{xi19} modeled
the circumstellar environment inferred mass loss rates $\lesssim$ 10$^{-4}$ M$_{\sun}$ yr$^{-1}$ which, while high, are broadly consistent with the rates which come out of the red supergiant phase in our models.

Next, we consider Kes~73. Kes~73 has a similar size and age as 1E 0102-7219, but is otherwise quite different. As seen in Table~\ref{tab:snrprop}, we measure a low sulfur line centroid and luminosity. As seen in Figure~\ref{fig:s_prop_gmm}, its spectral and dynamical properties straddle a boundary between lower mass progenitor models in a dense wind, and higher mass models in a low density wind. Based on an analysis of a deep {\it Chandra} observation, \citet{borkowski17} argued for a low mass progenitor (M$_{\mathrm{ZAMS}}$ $\lesssim$ 20~M$_{\sun}$) which resulted in a "fairly typical" IIP supernova. In contrast, \citet{chevalier05} compared the properties of the remnant to self-similar models, and inferred a more massive progenitor which would result in a IIL/b SN. Our models suggest a progenitor mass of $\gtrsim$ 22~M$_{\odot}$, which is in agreement that the remnant is from a IIL supernova.

Finally, we consider the case of G292.0+1.8 (hereafter, G292). Located at a distance of $\gtrsim$ 6 kpc \citep{gaensler03}, G292 has an angular size of $\sim$ 9$\arcmin$, corresponding to a physical diameter of 16 pc \citep{park07}. Based on the
proper motion of fast moving optical knots, \citet{ghavamian05} estimated its age to
be $\sim$ 3000 yr, consistent with the spin-down age of the central pulsar \citep[2900 yr;][]{camilo02}. Based on the X-ray properties of the shocked circumstellar material,
\citet{jjlee10} estimated a mass loss rate of the progenitor of $\dot{M}$ $\approx$ 2--5 $\times$ 10$^{-5}$ M$_{\sun}$ yr$^{-1}$, for a stellar wind speed of 10 km s$^{-1}$.

Recently, \citet{bhalerao19} performed an in-depth study of a $0.5$ Megasecond {\it Chandra}
ACIS-I observation of G292. Based on their detailed spectral analysis and estimates for
ejecta masses, they suggest that the progenitor mass of G292 lies within the range
of 13 M$_{\sun}$ $\lesssim$ $M$ $\lesssim$ 30 M$_{\sun}$. As seen in Table~\ref{tab:snrprop}, and seen in  Figures~\ref{fig:s_properties} (left) and ~\ref{fig:s_prop_gmm} (left), our models suggests a progenitor mass of $\gtrsim$ 20 M$_{\sun}$. Comparisons with the radius vs line centroid plots are consistent with this assertion. We therefore suggest that G292 is the remnant of a Type IIL SN, consistent with previous assert by \citet{chevalier05}.

One of the unanswered issues in \citet{pat15} was that while the small grid of models used in that initial study could reproduce the observed iron line centroids and luminosities seen in a large number of remnants, there was a poor relation in the dynamics between observed SNR radius and the spectral properties. Those authors asserted that this could possibly be addressed by investigating models with larger mass loss rates. They argued that the larger mass loss rate would not have a large impact on the dynamics, but would impact the luminosity and
other spectral qualities. As seen in this section, as well as in Figures~\ref{fig:slow-rad-fs} and ~\ref{fig:fast-rad-fs}, this hypothesis appears reasonable. 

\section{Application to Next Generation Observatories: Spatially Unresolved Remnants}
\label{unresolved}
Next-generation X-ray observatories such as a \textit{XRISM} and \textit{Athena} will have unprecedented spectral resolution compared to previous instruments. Such resolution will allow additional observables such as the S centroids mentioned above, or the He-like to H-like ratios of elements not examined here. While these instruments lack the spatial resolution necessary to resolve all but the largest and nearest remnants, their integrated spectra will carry important diagnostic information. For instance, Doppler shifts and absorption can be used to separate emission from different remnant regions and examine the spatial structure of the remnant. We defer discussion of these methods to a later publication, and choose to focus on what can be determined directly from the integrated spectra. 

\subsection{SNR Spectral Features for \textit{XRISM} and \textit{Athena}}
\label{nextgenspec}
\begin{figure}[htb]
    \centering
    \includegraphics[scale=0.23]{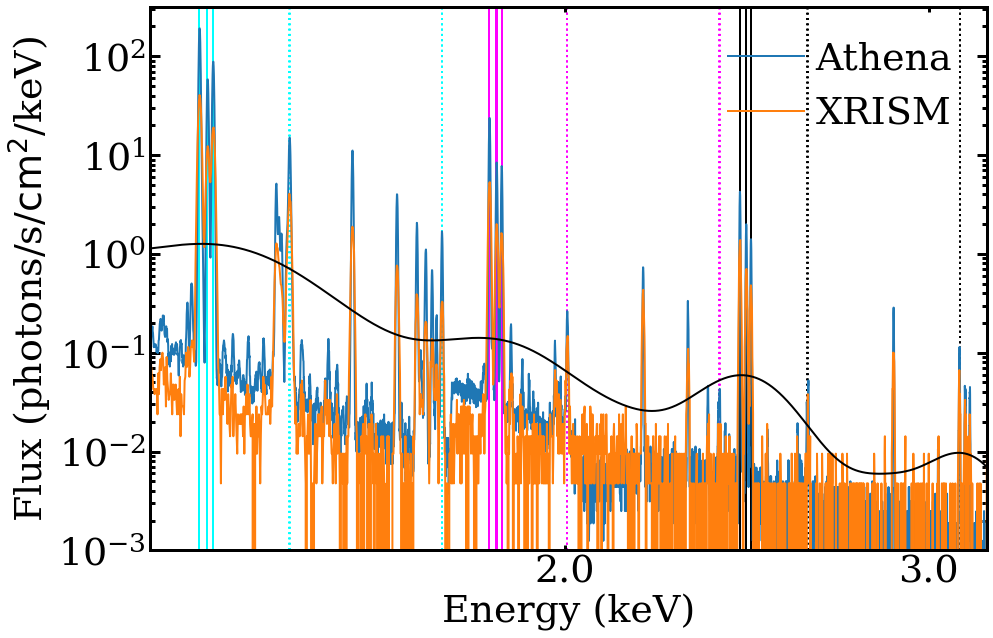}
    \caption{{\it Athena} (blue) and {\it XRISM} (orange) spectra plotted against a $100\  \rm eV$ resolution spectrum (black) for a $16.7 \rm M_{\odot}$ model in a slow wind normalized to the SMC for a 300 kilosecond exposure. The vertical lines indicate H-like (dashed), and He-like (solid) transitions for Mg (cyan), Si (magenta), and S (black).}
    \label{fig:midxray}
\end{figure}

\begin{figure}[htb]
    \centering
    \includegraphics[scale=0.23]{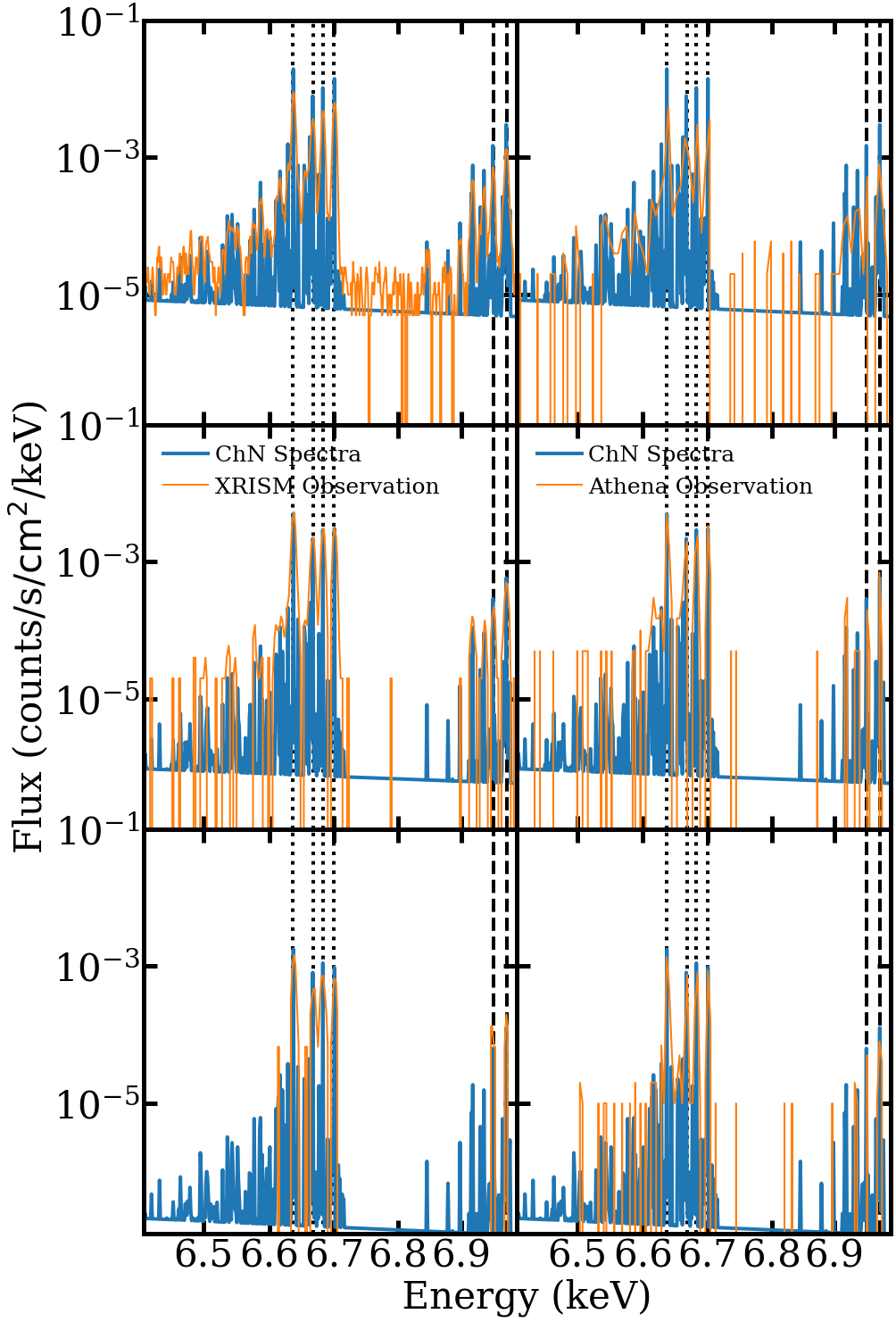}
    \caption{{\it XRISM} (left), and {\it Athena} spectra plotted against the original model spectra for the total shocked region of the three example models, normalized to the SMC. The XRISM spectra were exposed for $300 \rm ks$, and the Athena spectra were limited by the large number of counts at $\sim 1\ \rm keV$ and could only be exposed for $3 \rm ks$, $20 \rm ks$ and $300 \rm ks$ (top to bottom). The He-like and H-like lines of Fe are individually resolvable with both instruments.}
    \label{fig:fehires}
\end{figure}

\indent Next-generation X-ray telescopes will be able to observe entire SNRs in the local universe with observations of $\approx 10-100$ kiloseconds. Objects at distances of $\gtrsim 10 \rm kpc$ will not be resolvable to instruments such as \textit{Athena} or \textit{XRISM},  but the integrated spectra still contains a great deal of information. The resolving power of these new instruments allows for probing the He-like and H-like emission from lower-Z elements, as can be seen Figure~\ref{fig:midxray}. These simulated observations show what can be reasonably expected for a middle-aged SNR located within the Small Magellanic Cloud (SMC) for \textit{XRISM}, and \textit{Athena}.  He-like transitions are clearly resolvable for Mg, Si, and S, as are most H-like lines.  
He-like and H-like Fe emission still exhibit a model dependence, even for the example models presented above. Figure~\ref{fig:fehires}, presents side-by-side comparisons of \textit{XRISM} and \textit{Athena} for the three example remnants with a slow wind. Even with short exposure times, the He-like lines are still easily identified, and the H-like emission is discernible in most cases. 

\subsection{Helium-like Emission and Line Ratios}
\label{helikeem}
The spectral resolution available to next generation observatories will allow for the probing of emission lines from individual transitions. \textit{Hitomi} observations have already shown that it is possible to resolve the Forbidden (F,z), Intercombination ($\rm I_{1}$, x, and $\rm I_{2}$, y), and Resonance (R,w), transitions of He-like Fe \citep{hehitomi}. The ratios of these lines can be used to determine the underlying physical conditions present in the plasma \citep{porquet10}. The \textit{g}-ratio is defined as
\begin{equation}
    g\left(T_e \right) =\frac{z+x+y}{w}
\end{equation}
and has a dependence on the electron temperature. Additionally, the \textit{r}-ratio is defined as
\begin{equation}
    r\left(n_e \right) =\frac{x+y}{z}
\end{equation}
and has a dependence on the electron number density. 
\begin{figure*}[htb]
    \centering
    \includegraphics[scale=0.32]{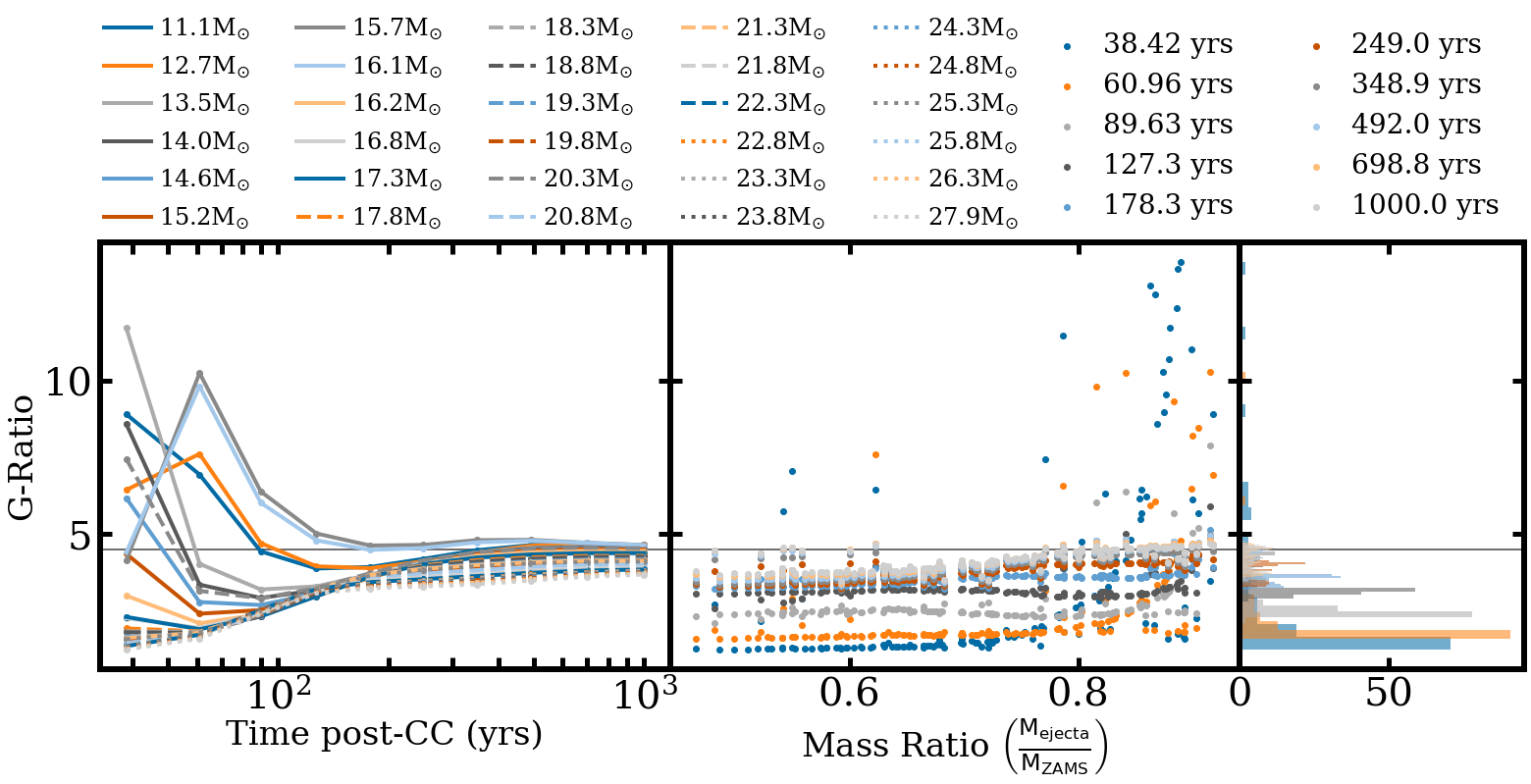}
    \caption{The Fe G-ratio extracted from the integrated spectra for the shocked region. In general, a larger ZAMS mass results in a lower intial G-ratio, with values eventually converging to $\approx 4.25$ (horizontal line). Lower mass models exhibit an early time peak, along with those models that experienced an early time blue loop $(15.7 \rm M_{\odot}, 16.1 \rm M_{\odot})$}
    \label{fig:g-fe}
\end{figure*} Unfortunately, there is not sufficient difference in the dependence on thermodynamic variables to differentiate between models, as they all represent similar NEI conditions. Additionally, the \textit{r}-ratio for high-Z elements requires significantly larger densities than are present in these SNR models. 

The \textit{g}-ratio does exhibit time and mass-dependent behavior, with the ratios of all models converging to $\approx 4$ within $\sim 10^3$ yrs. Additionally, models that exhibited a low mass loss rate (high mass ratio) often started with a higher initial \textit{g}, and exhibited a peak within the first $\sim10^2$ yrs, before settling near the limit. Objects where the mass loss was much lower than would usually be expected (ie. the blue loops) exhibited a lower initial \textit{g}-ratio, which peaks within $\sim 100~\rm yrs$ before quickly returning to the limit. Meanwhile, models that exhibited a higher mass loss rate started with a significantly lower \textit{g}-ratio, and rose to the limit. Figure~\ref{fig:g-fe} explores these behaviors in detail. There is a great deal of variability, especially at very early times when the total fraction of shocked material is still low, but when considering remnants younger than $\sim 100-200~\rm yrs$, He-like line ratios could be a useful probe of the mass loss rate, and the ZAMS mass of the progenitor. As in the forward shock only case, the line ratio begins by decreasing, reaching some minimum value before rising again.

\begin{figure}[htb]
    \centering
    \includegraphics[scale=0.16]{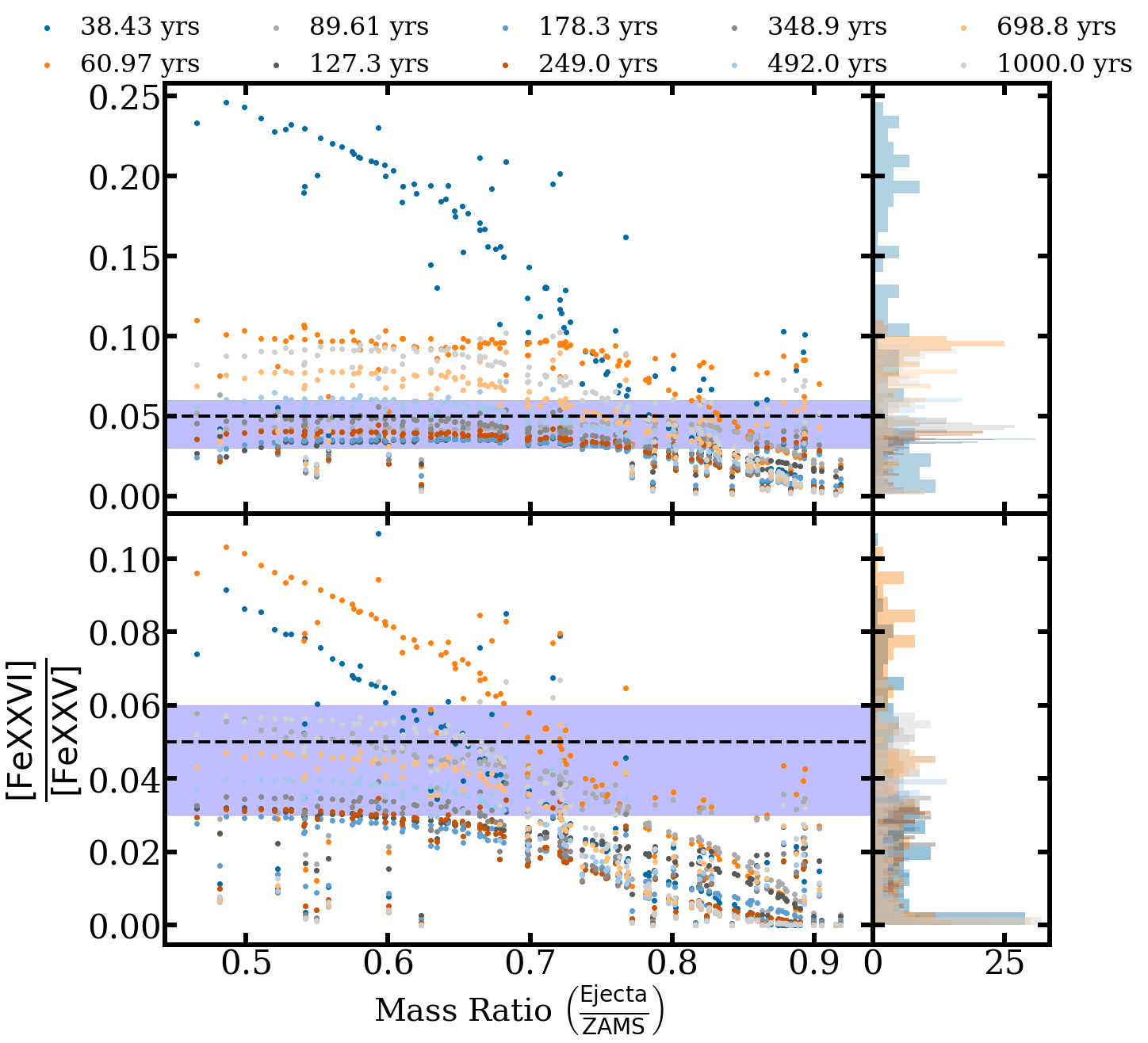}
    \caption{The $\frac{\rm \left[FeXXVI\right]}{\rm \left[FeXXV\right]}$ line ratios for the integrated shock region as a function of time and mass ratio for the slow wind (top) and fast wind (bottom). The early time behavior is largely the same, but there is a second rise at times ($\gtrsim 150~\rm yrs$).}
    \label{fig:int-shock-fe}
\end{figure}

The H-like to He-like line ratio behavior also changes slightly when considering emission from the entire shocked region. Unlike in the forward shock only case, the line ratio begins by decreasing, reaching some minimum value before rising again.  This is consistent with what is described in \citet{pat15}, where He- and H-like line emission is first dominated by shocked circumstellar material. But as the SNR evolves, shocked ejecta becomes more important to the integrated X-ray emission, causing it to rise, and altering the line ratios in ways that are dependent upon both the ejecta mass and circumstellar environment. Overall, the trend of line ratio vs. mass ratio still exists as in the FS case, with lower mass ratio models exhibiting a higher line ratio for a given remnant age. There also exists a flattening of the data below a mass ratio of $0.7$ in the fast wind, with those models maintaining a minimum line ratio of $\approx 0.03$ before rising again. The division appears more strongly in the fast wind case, and appears to be washed out by the increased density of the slow wind. These trends can be seen in Figure~\ref{fig:int-shock-fe}.

\section{Summary and Future Work}
\label{summary}
\indent Supernova remnant evolution is dependent on the mass loss history of the progenitor object. When considering line-driven winds it is possible to discern between progenitor models based on the spectra of the resulting remnant. Ejecta masses tend to cluster around a narrow range, with the main difference in composition being the CO core. There is variation in mass loss based on the wind prescription used, which leads to variation in the CSM density, but the overall trends are similar within a given wind prescription, and show an identifiable trend based on progenitor mass. the Fe-K$\alpha$ centroid is a reasonable discriminant of the progenitor type (Ia or CC) and the ZAMS mass. The He-like to H-like line ratios are a good proxy for the ionization state of the remnant. 

In cases where the stellar mass loss is well described by an isotropic wind, remnant information can be reliably recovered using the Fe-K$\alpha$ centroid and the radius. Without examining the individual composition and metallicity of a remnant, it is still possible to examine some aspects of the X-ray spectra and make inferences about the mass range, as in the case of G292.0+1.8 and others. Future X-ray missions will allow for additional observables such as the He-like line ratios which can be used to probe the CSM density and progenitor mass even in the case of a spatially unresolved remnant.

Models that have a more complex mass loss history, such as the blue-loop models will lead to a more complex CSM, as the lower mass loss rate during the loop will lead to a low density shell embedded within the denser RSG wind. That aspect is currently neglected, along with the high speed winds expected from YSG, BSG, and Wolf-Rayet stars. A more detailed investigation of those remnants is warranted. The current mass loss prescriptions also led to partially stripped envelope SNe only for higher mass models ($M\gtrsim22\rm M_{\odot}$). This is broadly consistent with the progression of mass stripping in supernova progenitors when considering IIP vs IIL SNe. For lower mass stars, additional mass loss prescriptions are required to remove the hydrogen envelope such as wave-driven mass loss or binary interactions.

Finally, Projection effects combined with shock Doppler boosting and intra-remnant absorption can be used to extract additional information from spatially unresolved remnant spectra. Modeling of these effects is currently being incorporated into the above framework, and will be the subject of a future work.

\begin{acknowledgements}
\section*{Acknowledgements}
\label{acknowledgements}
TEJ acknowledges support from NASA Astrophysical Theory Program \#80NSSC18K0566. He additionally acknowledges support from the Chandra X-ray Center, which is operated by the Smithsonian Institution under NASA contract NAS8-03060. S.H.L.
is supported by JSPS KAKENHI grant No. JP19K03913, and by the World Premier International Research Center Initiative (WPI), MEXT, Japan. D.M. acknowledges National Science Foundation support from grants PHY-1914448 and AST-2037297. This work was partially supported by JSPS Grants-in-Aid for Scientific Research “KAKENHI” (A: Grant Number JP19H00693). SN wishes to acknowledge support from the RIKEN Program of Interdisciplinary Theoretical and Mathematical Sciences (iTHEMS) \& the RIKEN Pioneering Program for Evolution of Matter in the Universe (r-EMU). Some of the computations in this paper were conducted on the Smithsonian High Performance Cluster (SI/HPC), Smithsonian Institution (\url{https://doi.org/10.25572/SIHPC}).
This work made use of the following software packages: MatPlotLib \citep{Hunter:2007}, NASA ADS, NumPy \citep{van2011numpy}, Pandas \citep{pandas}, astroML \citep{astroML}.
\end{acknowledgements}
\section*{Data Availability}

The data sets generated for this article will be available in the gitlab group pre-ccsne-wind at \url{https://gitlab.com/pre-ccsne-wind}. 
\bibliography{bibliography}
\bibliographystyle{aasjournal}

\end{document}